Federal Center for Technological Education of Minas Gerais
Program in Mathematical and Computational Modeling

# Absolute Governance: A Framework for Synchronization and Certification of the Corporate Contractual State.

**Antonio Augusto Hoffert Cruz Pereira**

Supervisor: Prof. PhD. Adriano César Machado Pereira
UFMG

**Antonio Augusto Hoffert Cruz Pereira**

# Absolute Governance:
## A Framework for Synchronization and Certification of the Corporate Contractual State.

Dissertation presented to the Program in Mathematical and Computational Modeling at the Federal Center for Technological Education of Minas Gerais, as a final requirement for obtaining the title of Master in Mathematical and Computational Modeling.

Concentration Area: Mathematical and Computational Modeling.

Research line: Intelligent Systems.

Supervisor: Prof. PhD. Adriano César Machado Pereira
UFMG

*To the three Marys, that guided me.*

Antonio Hoffert

# Acknowledgements

I acknowledge the serenity and experience of my supervisor in conducting the revisions and improvement of this dissertation, the synergy of my business partner in the implementation of the proposed methodology and the support of my wife during the period of this work.

> “Trust, but verify.”
>
> — *Russian proverb*

# Abstract

This dissertation addresses the challenge of ensuring transactional integrity and reducing costs in corporate governance through blockchain technology. We propose an on-chain methodology for certifying, registering, and querying institutional transactional status. Our decentralized governance approach utilizes consensus mechanisms and smart contracts to automate and enforce business rules. The framework aims to reduce the transaction costs associated with contractual measurement reports and enhance overall transactional integrity. We provide a detailed exploration of how blockchain technology can be effectively harnessed to offer a robust solution to these challenges, setting the stage for our proposed solution and its potential impact on corporate governance. The application of the methodology resulted in as average of 2% overbilling reduction.

# List of figures

# List of tables

# List of abbreviations and acronyms

| | |
|---|---|
| AG | Absolute Governance |
| ANOVA | Analysis of Variance |
| API | Application Programming Interface |
| B2B | Business to Business |
| BPMN | Business Process Modeling Notation |
| CDBC | Central Bank Digital Currencies |
| DApp | Decentralized Application |
| DLT | Distributed Ledger Technology |
| ERP | Enterprise Resource Planning |
| ESG | Environmental, Social, and Governance |
| FINRA | Financial Industry Regulatory Authority |
| GDPR | General Data Protection Regulation |
| JSON | JavaScript Object Notation |
| NIST | National Institute of Standards and Technology |
| SAAS | Software as a Service |
| SC | Smart Contract |
| SHA | Secure Hash Algorithm |
| SLA | Service Level Agreement |
| TOC | Total Operation Cost |
| TX | Blockchain Transaction |
| UETA | Uniform Electronic Transactions Act |
| UI | User Interface |
| UML | Unified Modeling Language |
| USD | United States Dollar |
| UX | User Experience |

# List of symbols

| | |
|---|---|
| $O$ | Operational input data |
| $P$ | Financial output payable data |
| $f$ | Function representing service level agreement rules |

# Summary

# 1 Introduction

The advent of blockchain technology heralded a transformative paradigm for the global financial system. As a decentralized and tamper-proof data repository, the distributed communication protocol offers a solution to execute a high volume of transactions in environments where trust is essential and supplanted by the incentive structure of the distributed design.

Blockchain technology, originating from Nakamoto's seminal work (NAKAMOTO, 2008), is fundamentally a form of *Distributed Ledger Technology* (DLT) that employs a unique data structure to ensure decentralized, transparent, and secure data storage across various network nodes. Initially, the architecture of blockchain technology was harnessed to support Bitcoin, the pioneering cryptocurrency. However, since its inception, blockchain applications have extended beyond the cryptocurrency domain, demonstrating immense utility in economic sectors (TAPSCOTT; TAPSCOTT, 2016).

A salient attribute of blockchains is their append-only nature, suggesting that while new blocks can be appended to the chain end, altering the content of an existing block requires modifying all subsequent blocks (CROSBY et al., 2016). This distinctive feature bolsters the security and integrity of data embedded in the blockchain.

At the heart of blockchain technology lies the concept of decentralized consensus, an attribute eliminating the need for a controlling entity or third-party trust to validate transactions. This consensus mechanism ensures unanimous agreement on the ledger's state among all network participants. As a result, the decentralized nature of the blockchain curtails potential fraud, substantially reducing transaction costs (TAPSCOTT; TAPSCOTT, 2016) and contributing to the institutional ESG score.

Nevertheless, significant challenges in utilizing blockchain for financial operations include establishing the credibility and truthfulness of financial data before logging into the protocol and managing access control to sensitive information. These challenges are especially pronounced in corporate financial data management, where disclosure constraints and data validation emerge as legal requirements and competitive advantages.

The focus of this dissertation is to propose a methodology to synchronize the transactional status of contracts between two companies through a private consensus mechanism applied to register measurement reports on a public blockchain. Within the methodology we suggest for recording corporate financial and inventory data, blockchain technology offers an ideal solution for creating a secure, transparent, and unchangeable transaction record. The dialectical process involves multiple parties with differing interests discussing interpretation contrasts before a network record is made, and coupled with blockchain's decentralization, it removes the need for a central authority. Moreover, the append-only nature of blockchain ensures that once financial or inventory data is recorded, it remains unaltered. Additionally, the inherent consensus mechanism of blockchain provides a robust validation method, promoting trust between institutions and encouraging more transactions.

Considering a contractual relation mutually dependent on the data of the transactional state, such as a service level agreement, the hypothesis studied is: Will the cost of synchronization of the transactional state between multiple organizations be compensated by the reduction on the total cost of operation *(TCO)*.

As a scientific proposition, we introduce a methodology termed *Absolute Governance* to foster contractual truthfulness, register, and abstract the physical-financial transactional data of both managerial reports and institutional inventories using blockchain technology. This structure deploys a multi-stakeholder consensus mechanism, obviating the need for a certifying authority or external audit vouching for its credibility.

## 1.1 The Centralized Governance Problem

Corporate governance is perceived as a complex system according to the classical definition of complexity theory. (GOERGEN et al., 2010) One of the most critical subsystems of governance is the information security methodology, a semantic ensemble that encompasses all the issues detailed henceforth.

The first challenge addresses the centralization of truth in institutional transactional records. In present-day accounting, operational and contractual records, the institution itself remains the sole source of transactional information. The absence of counter-parties with divergent interests to validate transactional state — where audits fail to meet this requirement due to a well-known dilemma termed the "agency problem" — paves the way for potential frauds, friction, and inefficiencies. (CHEN, 2022).

Another significant concern lies in the lack of version control, given the absolute mutability of accounting records over time.There's a void regarding cryptographic commitment (CRÉPEAU, 2011), which would immutably log the entire history of record modifications, inclusive of their final version. Logging such a history prohibits the deployment of *quantitative easing* techniques in account manipulation.

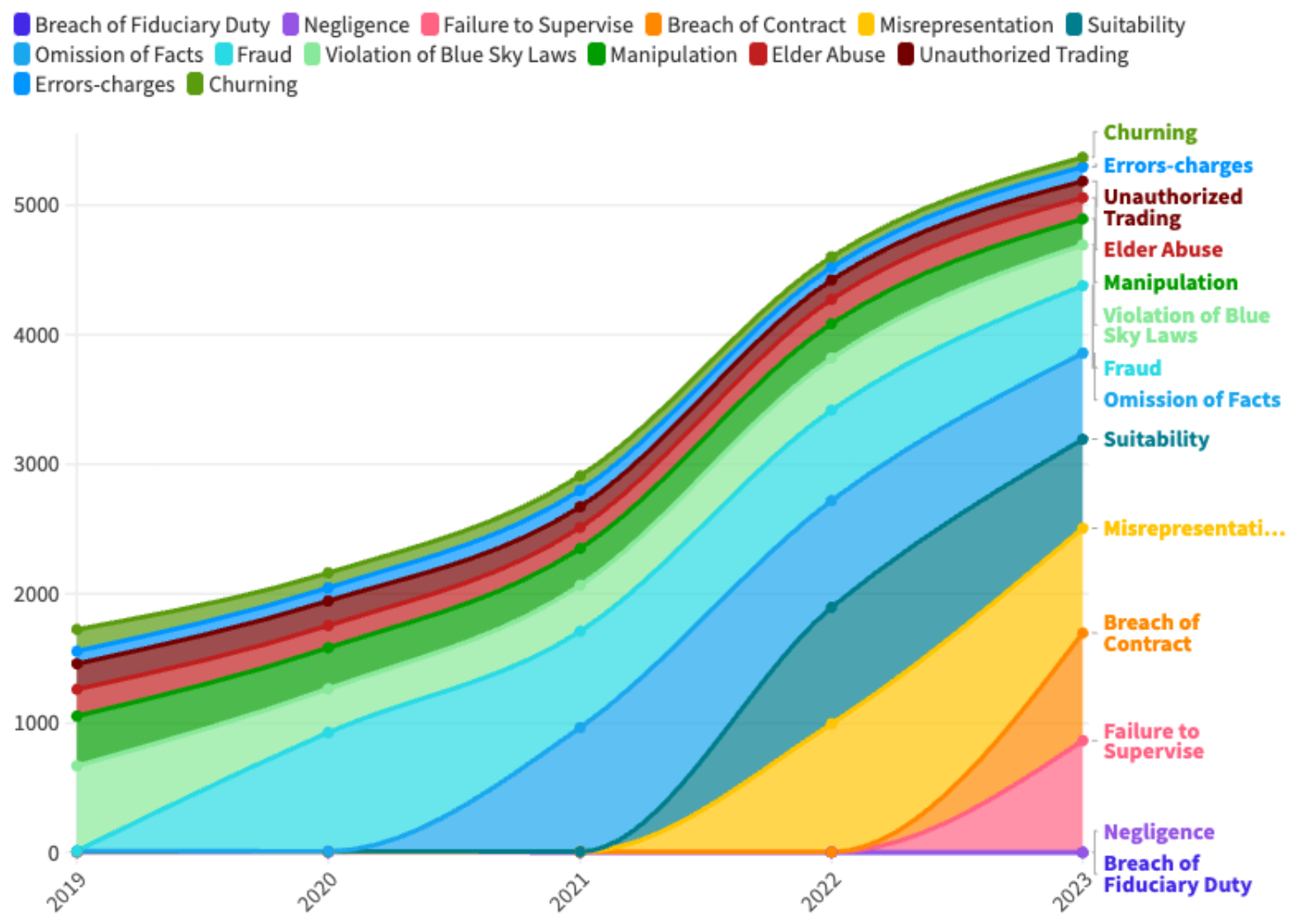

Figure 1 – Most frequent controversy types in contractual arbitration from 2019 to 2023.

Source: *FINRA, 2023.*

One of the most prominent contractual arbitration centers in the world is the Financial Industry Regulatory Authority that is receiving a crescendo of cases from 2019 to 2023. This tendency of demand for arbitrage illustrates the problematic towards centralized governed contracts.

Considering SLA — "Service Level Agreement" contracts between two institutions, centralized governance practices, regarding institutional knowledge management served organizations until the present moment in only two possible modalities. In the first one, considered primitive, one company, either supplier or customer, registers the operational measurements and provides a unilateral source of trust in which the other organization must passively accept. In the second approach, both institutions collected separated records of transactional data, stored on their respective ERP systems as a control mechanism that needs to be synchronized at each SLA billing cycle. This temporal and methodological hiatus in synchronizations leads to what can be formulated in a general sense as contractual disputes. There are several contractual controversy types, as all of them present financial impacts that result in a perceived underbilling or overbilling — these ambivalent omnipresent phenomena are not considered categories by arbitration research centers, as seen on Figure 1.1.

## 1.2 Motivation

The motivation for this dissertation is to reduce the transaction costs arising from uncertainties during the contractual cycle and the recording of ongoing operations between the parties involved. The objective is to stimulate a higher volume of economic transactions, improving trust between institutions by implementing a resilient and transparent mechanism to validate transactional states. The development of an Absolute Governance methodology aims to address the unsolved challenges of traditional governance by minimizing transactional uncertainties and fostering a verifiable environment conducive to economic transactions (DOE; ADAMS, 2021). By integrating blockchain into contractual governance, this study seeks to pioneer a new paradigm in corporate operations, offering a resilient method for data validation and privacy preservation in blockchain transactions (LEE; KIM, 2018) in order to establish Absolute Governance (AG) as a powerful tool for the industry 4.0 paradigm.

### 1.2.1 Anecdotal Reality

In 2019, the steel industry in Latin America faced one of its greatest challenges, a steel ladle leak that dispensed tons of molten steel into the metallic structure of the factory. After hundreds of millions in accounted losses, the CEOs of the steel industry and the raw material supplier that supported the equipment in question gathered in a regime consequence meeting to stipulate the responsibilities for the loss. A peculiarity in the case motivated this research: the document that was a setup requirement for the equipment to work on was not found after the incident. That was the pinnacle of the problem, and it raised only two possible scenarios. Either the document was never produced and there was a critical process failure, or it was kidnapped after the fact. This real anecdotal case illustrates the importance of registering the transactional state between companies in an immutable communication protocol. Otherwise, institutions may fall under the false impression of satisfactory information security, but when in contractual crises, it is discovered that centralized governance cannot withstand real-world entropy. Taking into account the teachings of that episode, the development of a Absolute Governance methodology began to be developed in 2019, rooted in real-world institutional necessities regarding the knowledge management of the contractual transactional state between companies.

## 1.3 Research Setting

The transaction state encompasses every patrimonial, financial, or operational change with which an institution can experiment, whether there are contracts, their relations overlap the transaction states of different institutions bonded by responsibilities and the expectation of remuneration (WACKER; YANG; SHEU, 2016; MAHER, 1997) This merged pathway defines the exact temporal and material realm needed to be synchronized between institutions in order to guarantee reciprocal fulfillment of a contract (WILLIAMSON, 2002; HART; MOORE, 1988). This synchronization is not only critical to managing and mitigating the risks associated with contractual relationships, but it also plays a significant role in improving organizational efficiency and competitiveness by reducing transaction costs and fostering trust between parties (WACKER; YANG; SHEU, 2016).

In every contractual relationship, there is data generation and transference regarding transactional events. These events can represent inventory changes, financial flows, or services provided. Regardless of its nature, the transaction needs to be satisfactorily understood and recognized by the parties involved. The elements for fostering this business reciprocity in addition to the storage, availability, and preservation of transactional data (MAHER, 1997), are means of legitimizing the presented reality considering its origin, the responsibilities, involvement, conflict of interest of the proponent, and means of control and verification by the validator.(MAYER, 1995; COASE, 1937)

Within this specific, although widespread business challenge, the Absolute Governance methodology encompasses a broad scope of opportunities to be applied in order to assist the synchronization of transaction state between institutions. Nevertheless, as a research strategy of this dissertation, the case study and analysis will focus on only one business situation that can be defined, measured, and compared: Service Level Agreement (SLA) contracts.

### 1.3.1 Specification of Research Scope

A Service Level Agreement (SLA) is a formal document that defines the level of service expected from a service provider by a customer (LEWIS, 2009). It details the metrics by which the service is measured, as well as the remedies or penalties should agreed-upon service levels not be achieved (JAIN, 2001). Typically, SLAs are part of a broader contract between businesses and their clients or between different departments within an organization, ensuring that services are delivered at a specified quality, availability, and responsiveness (STURM; MORRIS; JANDER, 2000).

Let $O$ represent the operational input metrics . Let $f$ be a function representing the business rules applied to these metrics, which could include conditions on service levels, penalties for noncompliance, and bonuses for exceeding targets. The financial output or payable, $P$, under the SLA can then be represented as:

$$\underbrace{P}_{\substack{Payable\\Output}} = \underbrace{f}_{\substack{Contractual\\Rules}} \left( \underbrace{O}_{\substack{Operational\\Input}} \right)$$

In the classical example, if the business rule specifies a penalty for a service downtime below 99.9% uptime, the function $f$ could be defined to calculate penalties based on the actual uptime percentage, $U$, as follows:

$$\text{Considering:} \quad \overbrace{U}^{\substack{(Uptime)\\Operational\\Input}} \qquad \underbrace{P}_{\substack{Payable\\Output}} = \underbrace{\begin{cases} P & \text{if } U \geq 99.9\% \\ P - C \cdot (99.9\% - U) & \text{if } U < 99.9\% \end{cases}}_{\substack{Contractual\\Rules}} \tag{1}$$

Here, $C$ represents the cost penalty for each percentage point (or part thereof) below the agreed uptime. However, according to the general formula of the SLA, considering that every payable amount that depends on an operational delivery falls under this category, it is reasonable to observe that the great proportion, if not the large majority, of Business-to-Business contracts are SLAs . In industries such as construction, utilities, aerospace, defense, food, and manufacturing, among others, SLAs receive different nomenclatures.

These performance or calibration contracts are pivotal for ensuring projects and services meet predefined standards. Research indicates that the aforementioned sectors allocate or derive over 90% of their revenues from this type of contracts, highlighting the critical role of contract management in improving vendor performance and optimizing costs. This justifies the definition of the scope of research of this dissertation for the potential for value generation through effective contract governance (COMPANY, 2020).

## 1.4 Core Objective

The objective of this research is to establish a blockchain based solution for the classical problem of contractual synchronization between institutions. This solution consists on an application implemented based on the Absolute Governance methodology of transparency, room for contradiction and auditability. This dissertation aims to explain how an application can incorporate a truth seeker methodology via versioning of contractual measurement reports and automation of business rules in order to promote a stable consensus between multiple parties regarding the execution, partial delivery or lack of execution, in an agreement.

### 1.4.1 Specific Objectives

#### 1.4.1.1 Construction of the Absolute Governance methodology:

Create a clear step-by-step roadmap towards the implementation of the Absolute Governance methodology to ensure the authenticity and integrity of both measurement reports and contractual rule's automation.

#### 1.4.1.2 Definition of the consensus mechanism architecture:

Outline the structural design of the consensus mechanism within the AG methodology. Illustrate the process that enables diffuse parties to agree on a single version of the truth by iteration.

#### 1.4.1.3 Presentation of the consequences of decentralizing and automating the contractual business rules in smart contract's quantitative algorithms:

Evaluate the consequences of transferring contractual governance to smart contracts by automating business rules. Answering the following research questions: Do companies usually follow their SLAs from an objective stance? Are there surplus benefits besides the direct economy of human hours in contract back-office and regulation?

#### 1.4.1.4 Instantiation of the methodology in a case study with implementation of the solution in real environments:

Perform comparative analysis to assess transactional efficiency in terms of costs, time, and resources between the current centralized governance system and the proposed methodology.

### 1.4.2 Specific Contribution

*"Much has been heard about the benefits of blockchain technology for corporations."* is an anecdotal proverb that denotes the lack of consolidated business applications in the marketplace amidst the plethora of theoretical works. The specific contribution of this dissertation to the academic scene on the topic is to provide the business and technical building blocks for applications to generate tangible value by improving contractual and overall governance and efficiency in corporations.

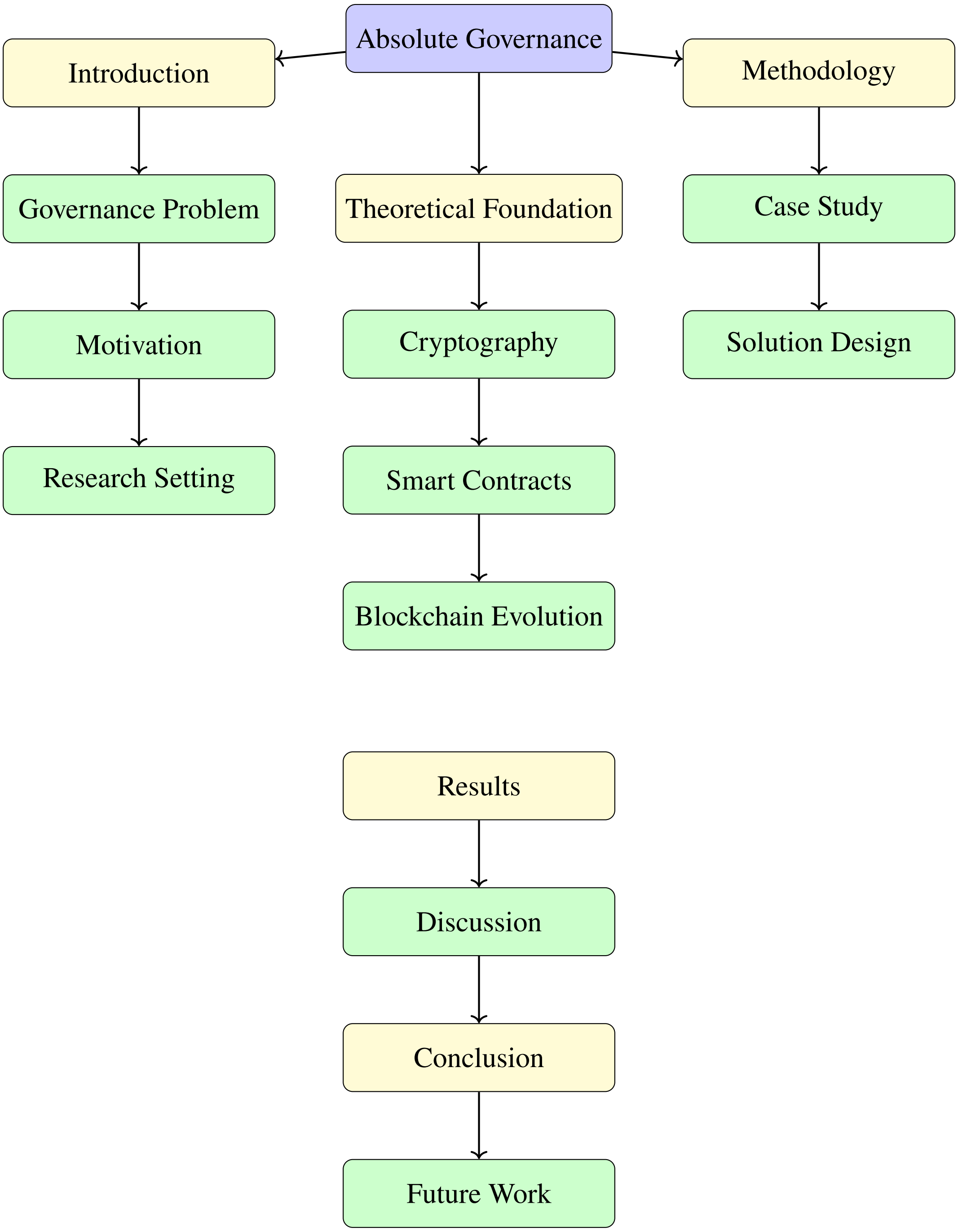

Figure 2 – Infographic of Absolute Governance Summary

Source: *Elaborated by the Author, 2024.*

# 2 Theoretical Foundation

This chapter builds on the introductory concepts presented in Chapter 1 and sets the technical groundwork for the proposed Absolute Governance methodology discussed in Chapter 4. It explores the multifaceted components of blockchain technology—smart contracts, cryptography, consensus mechanisms—and their roles in fostering a decentralized, secure, and efficient ecosystem.

Blockchain technology, fundamentally a decentralized ledger, has evolved far beyond its initial use in cryptocurrencies like Bitcoin. It now supports a plethora of decentralized applications (DApps) through smart contracts, primarily on platforms such as Ethereum. This chapter delves into the essential components of blockchain technology—smart contracts, cryptography, consensus mechanisms—and their pivotal roles in fostering a decentralized, secure, and efficient digital ecosystem.

## 2.1 Cryptography Fundamentals applied to Blockchain

Cryptography, the art and science of securing communication through the use of mathematical protocols and cypher algorithms, plays a pivotal role in safeguarding the integrity, confidentiality, and authenticity of digital interactions within blockchain networks. It is foundational for ensuring transaction security, maintaining participant anonymity, and achieving consensus across decentralized systems.

### 2.1.1 Objectives of Cryptography

Cryptography have defined objectives, each with a different purpose on communication, at the same time cryptography can be used to hide a message from observers it can be applied reversely to guarantee the authenticity of a signature:

**Authentication** Ensuring the transaction originates from a verified source and verifying the identity of participants (MENEZES; OORSCHOT; VANSTONE, 1996).

**Confidentiality** Protecting sensitive transaction data from unauthorized access and visibility (STALLINGS, 2017).

**Integrity** Safeguarding transaction data from unauthorized alterations during transmission (STINSON, 2005).

**Access Control** Managing who can view or alter information within the blockchain network (PERRIG et al., 2004).

**Non-Repudiation** Preventing any denial of a transaction by involved parties, typically enforced through digital signatures (KATZ; LINDELL, 2014).

### 2.1.2 Foundational Cryptographic Techniques in Blockchain

The primary cryptographic methods used in blockchain are encryption, decryption and hashing, before these concepts are explored a basic glossary needs to be presented.

#### 2.1.2.1 Terminology Used in Cryptography

**Plaintext** The original readable message or data that is fed into the encryption algorithm (SINGH, 1999).

**Ciphertext** The transformed, unreadable output of the encryption process, which conceals the plaintext's content during transmission (PAAR; PELZL, 2010).

**Encryption** The process of converting plaintext into ciphertext through the application of a cryptographic algorithm and a key (SCHNEIER, 1996).

**Decryption** The inverse process of encryption; it involves converting ciphertext back to plaintext using a cryptographic key (SMART, 2016).

**Key** A numeric or alphanumeric string utilized in the encryption and decryption processes (DIFFIE; HELLMAN, 1976).

**Private Key** An exclusive key accessible only to the signer, used for generating digital signatures (RIVEST; SHAMIR; ADLEMAN, 1978).

**Public Key** A key that is publicly accessible and used for verifying digital signatures (ELGAMAL, 1985).

#### 2.1.2.2 Symmetric-Key Cryptography

Also known as secret-key cryptography, this technique uses a single key for both encryption and decryption processes. The primary advantage of symmetric-key cryptography is its efficiency; operations are generally faster and less computationally intensive compared to asymmetric cryptography. This makes it suitable for scenarios where large volumes of data require encryption, such as file storage or database encryption. In symmetric-key cryptography, the same key (secret key) is used to encrypt and decrypt a message. When a message is sent, the sender uses the secret key to encrypt the data. Upon receiving the encrypted data, the recipient uses the identical secret key to decrypt the message, restoring it to its original form.

The major challenge in symmetric-key systems is the secure distribution of the secret key. Since both sender and receiver must possess the same key, and it must remain secret, robust mechanisms must be in place to prevent unauthorized access or interception during key exchange. Symmetric-key algorithms like AES (Advanced Encryption Standard), DES (Data Encryption Standard), and 3DES are widely used in sectors where data throughput and encryption speed are critical, such as banking, streaming services, and secure file storage (BELLARE; ROGAWAY, 1997).

#### 2.1.2.3 Asymmetric-Key Cryptography

Employs a public-private key pair to enhance security; this method is fundamental to modern digital communication and security, particularly in digital signature applications and encryption over open networks. What one key does only the other reverts. The public key can be distributed openly without compromising the security of the private key, which remains confidential. For encryption, a sender will encrypt the message using the recipient's public key. Only the recipient's corresponding private key can decrypt this message, ensuring that even if the message is intercepted, it cannot be deciphered without access to the private key.

Common asymmetric algorithms include RSA (Rivest–Shamir–Adleman), ECC (Elliptic Curve Cryptography), and DH (Diffie-Hellman). These algorithms are extensively used for secure data transmission, digital signatures in software distribution, secure email, and establishing secure connections between web servers and browsers (SSL/TLS) (DIFFIE; HELLMAN, 1976).

#### 2.1.2.4 Public and Private Key Generation

- **Private Key Generation:** A randomly generated secret number, denoted as $d$, used to sign transactions ensuring security and ownership.
- **Public Key Generation:** Computed from the private key, the public key $Q$ is given by $Q = dG$, where $G$ is a predefined point on the elliptic curve known as the generator point.

Elliptic Curve Cryptography (ECC), is the modern algorithm of choice for key-pair generation. ECC's strength lies in the complexity of the Elliptic Curve Discrete Logarithm Problem (ECDLP), making it computationally infeasible to derive the private key even if the public key is known. The equation of an elliptic curve over a finite field $\mathbb{F}_p$ is typically defined as:

$$y^2 = x^3 + ax + b$$

where $4a^3 + 27b^2 \neq 0$ to ensure the curve is non-singular. This condition ensures the curve does not contain any singular points (i.e., points where the curve touches or crosses itself), which is crucial for maintaining well-defined, invertible operations necessary for cryptographic applications. One of the primary reasons for the strength of ECC is the Elliptic Curve Discrete Logarithm Problem (ECDLP). This problem involves finding an integer $k$, given an elliptic curve point $G$ (the generator point) and another point $Q$ on the curve, such that $Q = kG$. The difficulty of solving the ECDLP grows with the size of $k$, providing a secure foundation for cryptographic operations since the problem is computationally hard. The operation of point multiplication, which is the calculation of $kG$ on an elliptic curve, exemplifies a cryptographic trapdoor function. This operation is straightforward to perform but extremely difficult to reverse, making it a robust choice for cryptographic security. Additionally, the security level of Elliptic Curve Cryptography can be finely adjusted by modifying the size of the field and the curve parameters. This flexibility allows the security level to be scaled up to counteract future increases in computing capabilities, including potential threats posed by quantum computing technologies.

#### 2.1.2.5 Signing with Private Keys

In blockchain applications that utilize Elliptic Curve Cryptography (ECC), the digital signature process is fundamental to securing transactions and establishing trust. This process begins with the generation of key pairs. A private key, denoted as $d$, is randomly chosen from the set $\{1, 2, \ldots, n-1\}$, where $n$ is the order of the curve. The public key $Q$ is subsequently derived by calculating $Q = dG$, with $G$ representing the generator point on the elliptic curve. This point is crucial for the cryptographic strength of the ECC and is known for its properties that facilitate secure operations.

Once the key pair is established, signing transactions involves creating a unique cryptographic signature that asserts the integrity and authenticity of the transaction data. The process starts by computing a cryptographic hash of the transaction data using SHA-256, resulting in a hash that acts as a compact representation of the data. This hash is then signed with the private key to produce a signature pair $(r, s)$, calculated as follows:

$$r = (x_1 \mod n), \quad s = k^{-1}(z + rd) \mod n,$$

where $k$ is a carefully selected random nonce that must remain confidential and $z$ is the hash of the transaction. The choice of $k$ and the secure computation of $s$ are critical to ensuring the non-repudiation and integrity of the signature. The security of a private key relies on the probability of successful attack that largely depends on the size of the key space, which is exponentially related to the length of the key:

$$\text{Key Space Size} = 2^{\text{Key Length}}$$

For example, a 256-bit key offers a vast key space of $2^{256}$ possible combinations. The immense size of this key space makes brute force attacks impractical with current computing technology, as the time and computational power required to test all possible keys would be prohibitive.

#### 2.1.2.6 Wallet Address Generation

Generating a wallet address from a public key involves several cryptographic transformations to ensure security and ease of use. The public key $Q$ undergoes a two-step hashing process, first with SHA-256 and then with RIPEMD-160, which not only secures the public key but also reduces its size for practical use. Following this, a network-specific byte is prefixed to enhance security and differentiation between different networks, and a checksum is generated by hashing the result twice with SHA-256. This checksum is critical as it verifies the integrity of the address. The full sequence is then encoded using Base58Check, a method that provides a compact, error-check resistant representation of the address suitable for easy handling in the blockchain ecosystem.

### 2.1.3 Cryptographic Hash Functions

Because of the basilar utility derived from hash functions in digital applications, this subsection will be dedicated to its high-level clarification. Hash functions process input data to produce a fixed-size string, typically for integrity checks and authentication purposes. Despite many potential inputs, hash functions are designed to make finding two different inputs that produce the same output (collision) computationally infeasible, thus enhancing security.

**Message Integrity Check (MIC)** Sends a hash of the message to verify integrity, with the MIC typically encrypted for security (BELLARE; CANETTI; KRAWCZYK, 1996).

**Message Authentication Code (MAC)** Involves sending a keyed hash of the message to authenticate the sender and ensure message integrity, often alongside optional message encryption (KRAWCZYK; BELLARE; CANETTI, 1997).

**Digital Signature** Uses encryption with the sender's private key to provide a secure and verifiable means of confirming the origin and integrity of a message (JOHNSON; MENEZES; VANSTONE, 2001).

Hash functions are mathematical algorithms that take an input (or 'message') and return a fixed-size string of bytes. The output, usually referred to as the hash, is typically a fixed-size bit string (the digest), which appears random. Hash functions derive their mathematical strength and security from their properties, which make them a tool of choice in cryptography. The strength of hash functions mathematically derives from the properties they satisfy:

- **Pre-image resistance** – Given a hash output, it should be computationally infeasible to reverse it to any original input that hashes to that output.

- **Second pre-image resistance** – Given an input and its hash, it should be computationally infeasible to find another input that has the same hash.
- **Collision resistance** – It should be computationally infeasible to find two different inputs that produce the same output.

These properties ensure that hash functions are practically irreversible, making them excellent unilateral cryptographic references (ROGAWAY; SHRIMPTON, 2004).

#### 2.1.3.1 Hash Collisions

A hash collision occurs when two distinct inputs produce the same hash output. Despite the computational unfeasibility of finding collisions due to the collision resistance property, they are theoretically possible due to the infinite number of possible inputs and finite number of outputs. Advanced hash functions are designed to minimize the probability of collisions, which is crucial for maintaining the integrity and security of data (BELLARE; CANETTI; KRAWCZYK, 1996). To understand how secure a hash is, the combinatorial birthday paradox must be mentioned.

##### 2.1.3.1.1 Understanding the Birthday Paradox in Cryptography

The Birthday Paradox refers to the counterintuitive probability problem, which queries the minimum number of people required in a room for there to be a better than even chance that at least two of them share the same birthday. In a non-cryptographic context, it surprisingly only requires 23 people to reach a probability greater than 50% that two individuals share the same birthday, given there are 365 days in a year. Mathematically, the probability $P(n)$ that at least two people out of $n$ share the same birthday is calculated as:

$$P(n) = 1 - \prod_{k=0}^{n-1} \left(1 - \frac{k}{365}\right)$$

This formula represents the complement of the probability that all $n$ birthdays are different. The relevance of the Birthday Paradox to hash functions can be seen when assessing collision resistance. The paradox suggests that for a hash function with a $b$-bit output, the expected number of random inputs needed to find a collision is not $2^b$ but rather $2^{b/2}$ due to the square root reduction implied by the paradox. This is significant in security because it means that finding a collision is computationally more feasible than it might initially seem if one were considering direct probabilities only. For a hash function producing a 256-bit output, the naive expectation might be that $2^{256}$ different inputs are needed for a collision. However, due to the Birthday Paradox, only about $2^{128}$ (i.e., the square root of $2^{256}$) are required to find a collision with high probability. Assuming a scenario where a highly specialized network can compute $10^{12}$ (one trillion) hashes per second, the total required hashes for a probable collision would be: $2^{128} \approx 3.4 \times 10^{38}$ hashes. In order to find the time needed, divide the total hashes required by the hash rate per second:

$$\text{Time (seconds)} = \frac{2^{128}}{10^{12}} \approx \frac{3.4 \times 10^{38}}{10^{12}} = 3.4 \times 10^{26} \text{ seconds}$$

Converting this to years:

$$\text{Time (years)} = \frac{3.4 \times 10^{26} \text{ seconds}}{60 \times 60 \times 24 \times 365.25} \approx 1.08 \times 10^{18} \text{ years}$$

To put that in perspective, it is about 78 billion times longer than the estimated age of the universe, which is about 13.8 billion years.

### 2.1.4 Applications of Hash Functions

Cryptographic hash functions have a wide range of applications in information security:

**Data Integrity Verification** Hashes are used to ensure that data has not been altered, intentionally or accidentally, from its original form. By computing the data's hash at storage and then re-computing it at retrieval, any change in the hash indicates a change in the data, hence a breach in integrity.

**Password Storage** Storing passwords as hashes in the database ensures that even if the database is compromised, the actual passwords are not exposed. Only the hash of the password is stored, and during authentication, the password provided is hashed and compared with the stored hash.

**Digital Signatures** Digital signatures play a vital role in ensuring the authenticity and integrity of a document.

The understanding of the process of a digital signature is a prerequisite to introduce the Absolute Governance Methodology later in this chapter.

### 2.1.5 Digital Signatures and Their Integral Role in Blockchain Security

Digital signatures deploy multiple critical functions, from transaction authentication to the maintenance of integrity across the network. Here is how the process of digitally signing works step-by-step:

1. **Document Hashing** Initially, a hash function is applied to all the data within a document. Even a small change in the document, such as altering a comma, will produce a significantly different hash (this property is known as the avalanche effect).
2. **Signing** The document's hash is then encrypted with the signer's private key. This encrypted hash, along with the hashing algorithm, constitutes the digital signature.
3. **Transmission** The original document is sent with the digital signature.
4. **Verification** The recipient decrypts the signature using the signer's public key, revealing the hash. The recipient then hashes the received document using the same hash function and compares this hash to the decrypted hash. If they match, it confirms the document's integrity and authenticity.

This process ensures that the document is exactly as it was when signed by the holder of the private key, and any tampering would invalidate the hash comparison (JOHNSON; MENEZES; VANSTONE, 2001).

**Digital Signature Mechanism** Digital signatures utilize asymmetric cryptography, involving a public and a private key. The private key, which is kept secret, is used to generate a signature on a message, while the public key, which is distributed and accessible to other network participants, is used for verifying the authenticity of the signature. (MENEZES; VANSTONE; OORSCHOT, 2018; SILVERMAN, 2006).

**Authentication and Verification** Digital signatures ensure that messages or transactions originate from specific, verifiable sources, thereby confirming the sender's identity and preventing impersonation. The verification process involves checking whether the signature is valid under the public key and if the message has been altered post-signing (BONEH; LIPTON, 1999). Once data is signed, any alteration invalidates the signature, thereby safeguarding against tampering and fraud (KATZ; LINDELL, 2014).

**Non Repudiation** In addition to authentication and integrity, digital signatures provide non-repudiation, which prevents a party from denying the authenticity of their signature on a transaction, thereby binding them to the terms of the transaction as it was recorded in the blockchain. This aspect is particularly crucial in legal and financial applications where proof of commitments is required (ZIMMERMANN, 1995).

**Mathematical Validity** The validity of digital signatures is underpinned by the strength of the private key and hashing algorithm's generator (GOLDREICH, 2001).

Smart contracts automate transactions and other specific actions on the blockchain without human intervention. Digital signatures are critical in activating these contracts, as they verify the parties involved and authorize the contract's terms to execute based on the coded conditions (CHRISTIDIS; DEVETSIKIOTIS, 2016).

#### 2.1.5.1 Digital Signatures Legal Frameworks and Court Acceptance

In the United States, digital transactions and signatures, including those executed via blockchain, are legally recognized under the *Electronic Signatures in Global and National Commerce Act (ESIGN, 2000)* and the *Uniform Electronic Transactions Act (UETA, 1999)*. These laws affirm that electronic records and signatures carry the same weight and legal effect as traditional paper documents and handwritten signatures. The State of Washington has enacted laws that validate the use of blockchain for maintaining digital records and recognizing blockchain signatures as legal instruments via the State of Washington Blockchain Enabling Act.

The EU has standardized the legal framework concerning electronic signatures and transactions via the *Electronic Identification and Trust Services Regulation (eIDAS, 2016)*, ensuring that digital signatures are admissible as evidence in legal proceedings across member states.

In China, blockchain-based evidence is formally accepted in legal disputes, particularly in the realm of Internet courts. The Supreme People's Court of China recognizes the integrity of data managed on blockchain, citing its immutability and transparency. A notable case involved the Hangzhou Internet Court, which confirmed the admissibility of blockchain evidence in a copyright infringement case in June 2018 (HANGZHOU. . . , 2018).

## 2.2 Smart Contracts

Smart contracts facilitate automated execution within blockchain networks, with Solidity being the predominant programming language for their implementation on Ethereum. The Ethereum Virtual Machine (EVM) ensures secure and isolated execution of these contracts, enabling decentralized applications that operate autonomously without central control (BUTERIN, 2013; WOOD, 2014).

### 2.2.1 Smart Contracts: Technical Overview and Execution Mechanisms

Smart contracts execute agreements within blockchain networks automatically. Predominantly implemented in Solidity, they run on the Ethereum Virtual Machine (EVM), leveraging the blockchain's security, immutability, and decentralization. They activate when predefined conditions are met, ensuring reliability and efficiency without manual intervention. The EVM employs "gas", a fee mechanism that scales with contract complexity, to act as a computational resource gatekeeper. This automation not only cuts the need for traditional escrow services, thereby reducing transaction costs, but also enhances operational efficiency. Smart contracts have evolved from basic scripts to complex multi-functional agreements capable of executing a wide range of tasks, including financial transactions, asset management, and compliance automation. This evolution reflects blockchain's transition from cryptocurrency infrastructure to a broader technology with applications across various domains (BUTERIN, 2013).

#### 2.2.1.1 Design Principles of Smart Contracts

Design principles of smart contracts include modularity, reusability, and upgradability. These principles ensure that smart contracts are maintainable, secure, and can be integrated with other contracts and systems. Modularity allows for dividing contracts into distinct blocks that handle specific functionalities, enhancing clarity and manageability (CLACK; BAKSHI; BRAINE, 2016). Well-known libraries and patterns, such as OpenZeppelin, provide tested and community-vetted frameworks for smart contract development. These libraries include secure implementations of common functionalities, reducing the risk of vulnerabilities and bugs in smart contract code (WöHRER; ZDUN, 2018; DIKA, 2017).

### 2.2.2 Technical Constraints and Evolution

Although smart contracts offer enhanced transactional efficiency, their development requires precise, objective input parameters for execution, which can limit their use to straightforward, conditional operations. The technological and legal frameworks surrounding smart contracts are evolving to address challenges such as contract amendment, termination, and interaction with off-chain data sources via oracles. This progression points to a future with more complex, legally compliant smart contracts capable of wider applicability in digital transactions (SZABO, 1997; GRIGG, 2004).

Scalability remains a significant challenge for smart contracts on blockchain networks like Ethereum. The network's capacity to process transactions, particularly during high demand, affects the performance and cost-effectiveness of deploying smart contracts. Solutions such as sharding and layer-two scaling are being explored to address these issues (CROMAN et al., 2016; EYAL et al., 2016).

#### 2.2.2.1 Dispute Resolution and Smart Contracts

Dispute resolution mechanisms specifically tailored to smart contracts are emerging. These include decentralized arbitration services and on-chain dispute resolution frameworks that leverage blockchain technology to ensure transparent and immutable decision-making (RASKIN, 2017).

### 2.2.3 Applications Across Industries

Smart contracts have transformative applications across various sectors. In the financial sector, they enable automated securities trading and risk management. In supply chain management, they facilitate real-time tracking and automated compliance with trade agreements.

#### 2.2.3.1 Impact on Governance and Compliance

Smart contracts can significantly impact governance and regulatory compliance by providing transparent mechanisms for enforcing regulations and corporate governance standards. They can automate reporting and compliance processes, reducing administrative burdens and enhancing transparency (WERBACH, 2018).

Each subsection builds on the theoretical foundations laid out earlier, exploring both the potential and limitations of smart contracts. This comprehensive approach not only deepens the academic exploration within your dissertation but also provides a solid basis for discussing the application of these technologies in later chapters.

## 2.3 Evolution of Blockchain Technology

Blockchain technology has transitioned from its origins as the infrastructure for Bitcoin to a foundational technology promising to revolutionize various sectors by enabling decentralized transactions, smart contracts, and improved security and transparency. The progression from Blockchain 1.0, focusing on cryptocurrency, to Blockchain 2.0, introducing smart contracts, and towards Blockchain 3.0, highlights its potential to foster decentralized applications beyond financial transactions (SWAN, 2015; ZHENG et al., 2017).

## 2.4 Decentralized Applications (DApps)

The introduction of smart contracts has catalyzed the development of decentralized applications (DApps), which operate on a peer-to-peer network of computers running a blockchain. DApps are open-source, function autonomously, and lack a central point of control. They represent a new paradigm for application development, encompassing sectors such as finance, social media, gaming, and governance (WANG et al., 2019; DANIEL; GUIDA, 2019).

## 2.5 Business Process Management through DAOs

Decentralized Autonomous Organizations (DAOs) embody a revolutionary approach to Business Process Management (BPM) by utilizing blockchain's inherent features such as transparency, security, and automation. DAOs reflect the principles of decentralized governance, where business rules and processes are encoded into smart contracts on platforms like Ethereum, allowing for autonomous business operations without centralized control mechanisms (BECK et al., 2018; WRIGHT; FILIPPI, 2015).

### 2.5.1 Introduction to DAOs in BPM

DAOs revolutionize traditional business governance models by embedding organizational rules into smart contracts, ensuring that business operations are executed automatically and transparently. This transformation not only democratizes business governance but also significantly enhances operational efficiency and reduces managerial overhead (CHRISTIDIS; DEVETSIKIOTIS, 2016; DUAN; XIONG; BAI, 2019).

### 2.5.2 Automating Governance Processes

Smart contracts within DAOs automate governance processes, from decision-making to execution of business operations, based on predefined rules and conditions. This automation streamlines business processes, ensures consistency and efficiency, and minimizes the potential for human error (TAPSCOTT; TAPSCOTT, 2016; CATALINI; GANS, 2016).

#### 2.5.2.1 Reducing Managerial Overhead

By automating governance tasks, DAOs allow organizations to redirect resources towards strategic initiatives, enhancing innovation and competitiveness. Additionally, the transparency and immutability of blockchain minimize the risks of fraud and corruption, further reducing operational costs (MORKUNAS; PASCHEN; BOON, 2019; OMAR; RAHMAN, 2019).

#### 2.5.2.2 Improving Operational Efficiency

DAOs enable real-time monitoring and optimization of business processes. The ability to update smart contracts in response to changing business strategies or regulatory requirements ensures that operations remain efficient and compliant. Furthermore, the global reach of blockchain facilitates seamless collaboration, enhancing business agility (MORKUNAS; PASCHEN; BOON, 2019; OMAR; RAHMAN, 2019).

## 2.6 Applications Across Industries

This section examines the impact of smart contracts across various industries, focusing on three key points: process and information integrity, contract execution automation, and tracking and provenance. The following tables provide a detailed analysis of applications, addressing specific pain points for each industry.

| Industry | Pain Points | Applications | Value Generated |
|---|---|---|---|
| Financial Sector | 1. Transaction Integrity<br>2. Automated Contract Execution<br>3. Asset Tracking and Provenance | - Decentralized Finance (DeFi)<br>- Automated securities trading<br>- Risk management | - Increased efficiency<br>- Cost reduction<br>- Enhanced transparency |
| Supply Chain Management | 1. Data Integrity in Supply Chain<br>2. Automation of Compliance Processes<br>3. Real-time Tracking of Goods | - Real-time tracking<br>- Automated compliance<br>- Transparency and fraud reduction | - Improved efficiency<br>- Cost savings<br>- Enhanced trust |
| Healthcare | 1. Security of Patient Data<br>2. Automated Patient Consent<br>3. Tracking of Medical Supplies | - Patient data management<br>- Automated consent<br>- Supply chain transparency | - Enhanced data security<br>- Regulatory compliance<br>- Improved trust |
| Real Estate | 1. Integrity of Property Records<br>2. Automation of Property Transactions<br>3. Verification of Ownership and Title | - Smart property transactions<br>- Automated lease agreements<br>- Title verification | - Reduced fraud<br>- Faster transaction processes<br>- Lower transaction costs |
| Insurance | 1. Integrity of Claims Data<br>2. Automated Claims Processing<br>3. Fraud Detection and Prevention | - Automated claims processing<br>- Dynamic policy pricing<br>- Fraud detection and prevention | - Reduced operational costs<br>- Improved customer satisfaction<br>- Enhanced fraud detection capabilities |
| Logistics | 1. Data Integrity in Shipping<br>2. Automation of Shipping Contracts<br>3. Tracking of Cargo | - Automated shipping contracts<br>- Real-time cargo tracking<br>- Compliance with international trade | - Reduced shipping delays<br>- Lower operational costs<br>- Enhanced transparency in the supply chain |

Table 1 – Applications of Smart Contracts Across Various Industries

Source: *Elaborated by the Author, 2024.*

| Industry | Pain Points | Applications | Value Generated |
|---|---|---|---|
| Energy | 1. Transparency in Energy Transactions<br>2. Automation of Energy Distribution<br>3. Tracking of Energy Usage and Trading | - Peer-to-peer energy trading<br>- Automated energy distribution contracts<br>- Transparency in energy trading | - Increased efficiency<br>- Lower energy costs<br>- Improved transparency in energy transactions |
| Legal | 1. Integrity of Legal Documents<br>2. Automated Contract Management<br>3. Verification of Legal Documents | - Smart legal contracts<br>- Decentralized dispute resolution<br>- Automated verification of legal documents | - Reduced legal costs<br>- Faster dispute resolution<br>- Enhanced integrity and security of legal documents |
| Defense | 1. Data Security and Integrity<br>2. Automation of Defense Contracts<br>3. Tracking and Provenance of Equipment | - Secure communication protocols<br>- Automated procurement contracts<br>- Equipment lifecycle management | - Enhanced security<br>- Reduced procurement time<br>- Improved equipment tracking |
| Steel | 1. Quality Control Data Integrity<br>2. Automation of Supplier Contracts<br>3. Tracking of Raw Materials and Finished Goods | - Automated quality control reporting<br>- Supplier contract management<br>- Inventory tracking | - Improved quality assurance<br>- Reduced procurement costs<br>- Enhanced supply chain transparency |
| Oil & Gas | 1. Data Integrity in Resource Management<br>2. Automation of Extraction and Distribution Contracts<br>3. Tracking of Oil and Gas Shipments | - Automated contract execution<br>- Resource management<br>- Shipment tracking | - Enhanced resource management<br>- Reduced operational costs<br>- Improved shipment tracking |
| Government | 1. Integrity of Public Records<br>2. Automation of Public Services<br>3. Tracking and Provenance of Government Assets | - Public record management<br>- Automated service delivery<br>- Asset management | - Improved public trust<br>- Enhanced service efficiency<br>- Better asset management |
| Agriculture | 1. Integrity of Agricultural Data<br>2. Automation of Supply Contracts<br>3. Tracking of Crop and Livestock Provenance | - Farm-to-table tracking<br>- Automated supply contracts<br>- Quality assurance | - Improved food safety<br>- Reduced transaction costs<br>- Enhanced supply chain transparency |

Table 2 – Applications of Smart Contracts Across Various Industries

Source: *Elaborated by the Author, 2024.*

## 2.7 The Absolute Governance Methodology

## 2.8 Principles

The Absolute Governance is composed by three pillars, a software client/user application that follows the methodology should treat all actions of its users as signed transactions, all imputed information as hashes registered in the blockchain, and all shared decisions as a multi-signature consensus mechanism.

### 2.8.1 Principle 1: Authentication and Signature of Actions

**Definition:** All actions within the system, whether initiated by humans or machines, must be authenticated and digitally signed. This encompasses approvals, disapprovals, logins, access control, interactions with smart contracts, and process initiations.

**Implications:**

- **Identified Actor:** Each action must have a clearly identified actor responsible for the initiation of the action.
- **Object and Timestamp:** Alongside the actor, every action must be recorded with an associated object (the target or subject of the action) and a precise timestamp, ensuring that the sequence of events is unambiguously logged.

### 2.8.2 Principle 2: Integrity of Input Information

**Definition:** All information input into the system must be hashed and subsequently recorded on the blockchain. This process is essential for preserving the integrity and immutability of data.

**Implications:**

- **Data Integrity:** By hashing information before registration, the system safeguards against unauthorized modifications and verifies the authenticity of the data without revealing its contents.

- **Blockchain Registration:** Storing the hash on a blockchain ensures that any tampering with the data can be detected, as the ledger provides a secure and immutable record.

### 2.8.3 Principle 3: Consensus on Shared Data

**Definition:** Any shared data, including measurement reports, decisions, and business rules, must be agreed upon by all relevant parties through a consensus mechanism before it is accepted as official.

**Implications:**

- **Multiple Parties Agreement:** This principle mandates that all parties involved in the transaction or decision-making process must reach a consensus, which promotes fairness and mutual trust.
- **Officialization of Data:** Data only becomes official and actionable once consensus is achieved, ensuring that all parties acknowledge and accept the outcome of the process.

## 2.9 Advantages of an Organization operating by Absolute Governance

A business operating under a managerial application that implements the Absolute Governance methodology gains significant advantages:

- **Enhanced Security and Integrity**: By ensuring all actions are authenticated and signed, and all information is hashed and stored on the blockchain, businesses can guarantee data integrity and security, reducing the risk of tampering and fraud.
- **Transparency and Trust**: The immutability and transparency of blockchain records foster trust among stakeholders, as all transactions and decisions are verifiable and traceable.

- **Efficiency and Automation**: Automating processes through smart contracts reduces manual intervention, speeds up operations, and minimizes errors, leading to significant cost savings and increased operational efficiency.
- **Decentralized Decision-Making**: Consensus mechanisms ensure that all relevant parties agree on shared data and decisions, promoting fairness and mutual trust, and enhancing the democratic nature of business operations.
- **Regulatory Compliance**: Automated compliance processes and immutable records help businesses adhere to regulatory requirements more efficiently, reducing the risk of non-compliance and associated penalties.
- **Traceability**: Blockchain provides an immutable record of all transactions, enabling detailed traceability of assets and products throughout their lifecycle, which is crucial for industries like supply chain, agriculture, and pharmaceuticals.
- **Origination of New Business Models**: Decentralized applications enable the creation of new, digitally-oriented business models that were previously not feasible. Examples include parametric insurance products that automatically process claims based on predefined criteria and are regulated and executed entirely through smart contracts.

# 3 Related Work

This chapter explores the integration of blockchain technology in the management of Service Level Agreements (SLAs), highlighting significant advancements and identifying gaps in existing research. Recent studies have proposed various blockchain-based frameworks for SLA conformance validation, monitoring, and enforcement, underpinning the technology's potential to replace traditional trust mechanisms with a decentralized and immutable ledger.

## 3.1 Blockchain in General Governance

To contextualize this research within the existing body of knowledge, recent studies on the application of blockchain in governance and corporate structures offer diverse insights:

*"Using Blockchain for Global Governance: Past, Present and Future"* by Tiwari and Pal (2023) (TIWARI; PAL, 2023) provides a comprehensive examination of the research on blockchain governance and proposes a conceptual framework for its application in both public and private sectors globally. This study is both qualitative and quantitative, employing the Preferred Reporting Items for Systematic and Meta-Analysis (PRISMA) and bibliographic analysis using VOSviewer visualization tool and R Studio. The findings highlight the multidisciplinary interest in blockchain's role in governing nations and corporations, showcasing numerous instances where governments and corporations have implemented the technology in areas impacting the public and stakeholders. The study concludes that blockchain has significant potential to enhance governance frameworks by improving transparency, accountability, and efficiency.

The OECD paper *"The Potential for Blockchain Technology in Corporate Governance"* (Organisation for Economic Co-operation and Development, 2023) delves into the disruptive potential of blockchain in financial services and its subsequent impact on corporate governance. It discusses recent applications of blockchain in financial services, such as smart contracts and decentralized finance (DeFi), and examines regulatory responses to these innovations. The paper emphasizes the need for balanced regulatory frameworks that foster innovation while protecting stakeholders. It concludes that blockchain can significantly improve corporate governance by enhancing transparency, reducing fraud, and streamlining regulatory compliance.

*"Blockchain Technology for Corporate Governance and Shareholder Activism"* by Anne Lafarre and Christoph Van der Elst (LAFARRE; ELST, 2023) explores how blockchain can address inefficiencies in corporate governance, particularly in the relationship between shareholders and companies. The study focuses on restructuring the Annual General Meeting of Shareholders (AGM). It suggests that blockchain can enhance shareholder monitoring and corporate bonding by enabling real-time voting and transparent decision-making processes. The research concludes that blockchain can reduce agency costs, facilitate transparency and trust, and improve the speed and efficiency of shareholder decision-making, thereby strengthening corporate governance structures.

*"The Supply Chain Has No Clothes: Technology Adoption of Blockchain for Supply Chain Transparency"* by Kristoffer Francisco and Rodger David Swanson (FRANCISCO; SWANSON, 2023), published in January 2018 in "Logistics", examines how blockchain technology can enhance transparency in supply chains. The authors utilize the Unified Theory of Acceptance and Use of Technology (UTAUT) to analyze blockchain adoption in supply chain traceability. The study develops a conceptual model that suggests blockchain can significantly reduce transaction costs, increase transparency, and improve decision-making speed in supply chains. The research concludes with implications for supply chain management, emphasizing blockchain's potential to transform traditional supply chain operations by ensuring greater visibility and accountability.

## 3.2 Blockchain in Service Level Agreement Management

The integration of blockchain technology into Service Level Agreement (SLA) management represents a significant advancement in ensuring contract compliance and transparency. Recent research has explored various aspects of this integration.

Neidhardt et al. (NEIDHARDT; KöHLER; NüTTGENS, 2018) propose a blockchain-based framework for SLA conformance validation, emphasizing the technology's potential to replace traditional trust mechanisms with a decentralized, immutable ledger. Their research focuses on the technical architecture and operational procedures required to implement such a framework effectively. The study concludes that blockchain can provide a reliable and tamper-proof method for tracking SLA compliance, thus enhancing trust between service providers and clients.

Uriarte et al. (URIARTE; NICOLA; KRITIKOS, 2018) explore the use of smart contracts on the Ethereum blockchain to automate SLA monitoring and enforcement. They highlight the flexibility of Ethereum's platform but also point out limitations related to permissionless participation and cost predictability. The study presents a detailed analysis of how smart contracts can be designed to automatically monitor SLA parameters and execute predefined actions in case of violations. The authors conclude that while the technology shows promise, further research is needed to address scalability and cost issues.

Scheid et al. (SCHEID et al., 2019) demonstrate the practical application of Ethereum for SLA monitoring and enforcement through a prototype implementation. Their work highlights the benefits of using blockchain for automated and transparent SLA management but also identifies challenges such as transaction costs and the need for permissioned blockchain environments to enhance security and predictability.

Alzubaidi et al. (ALZUBAIDI et al., 2019) focus on the advantages of private blockchain networks for SLA management. They argue that private blockchains can provide a more controlled and predictable environment compared to public blockchains like Ethereum. Their research includes a case study demonstrating how private blockchain networks can be used to manage SLAs effectively, with a particular focus on reducing operational costs and improving data privacy.

## 3.3 Blockchain in Specific Applications

*"A Comprehensive Review of Blockchain Technology in Supply Chain Management"* by various authors(AUTHORS, 2023b) reviews the state-of-the-art applications of blockchain technology across multiple domains, including supply chain management. It discusses how blockchain can enhance transparency, traceability, and trust in supply chains by providing a secure and decentralized way to track transactions and data exchanges. The review identifies key benefits and challenges of blockchain adoption in supply chains, concluding that while the technology holds transformative potential, issues related to integration, scalability, and regulatory compliance need to be addressed.

| Study Name | Research Scope | Application | Contribution |
|---|---|---|---|
| Using Blockchain for Global Governance: Past, Present and Future | Blockchain governance research globally. | Public and private sector governance. | Conceptual framework for blockchain governance, highlighting transparency and efficiency improvements. |
| The Potential for Blockchain Technology in Corporate Governance (OECD) | Blockchain's impact on corporate governance. | Financial services, regulatory frameworks. | Enhances transparency, reduces fraud, and streamlines compliance. |
| Blockchain Technology for Corporate Governance and Shareholder Activism | Addressing inefficiencies in corporate governance. | Shareholder monitoring and AGMs. | Real-time voting, reducing agency costs, and improving decision-making efficiency. |
| The Supply Chain Has No Clothes: Technology Adoption of Blockchain for Supply Chain Transparency | Blockchain adoption in supply chain traceability. | Supply chain transparency, traceability. | Reduces transaction costs, increases transparency, and improves decision-making speed. |

Table 3 – Summary of Blockchain Studies in General Governance

Source: *Elaborated by the Author, 2024.*

| Study Name | Research Scope | Application | Contribution |
|---|---|---|---|
| Blockchain-Based SLA Validation Framework | Blockchain framework for SLA validation. | SLA conformance validation. | Reliable and tamper-proof tracking of SLA compliance. |
| Smart Contracts for SLA Monitoring | Smart contracts for SLA monitoring. | SLA monitoring and enforcement. | Automates SLA parameters, highlights scalability and cost issues. |
| Ethereum-Based SLA Monitoring Prototype | Ethereum prototype for SLA monitoring. | SLA monitoring and enforcement. | Demonstrates benefits and challenges of blockchain in SLA management. |
| Private Blockchain Networks for SLA Management | Private blockchains for SLA management. | Private blockchain networks for SLA. | Controlled environments, reduced operational costs, improved data privacy. |

Table 4 – Summary of Blockchain Studies in SLA Management

Source: *Elaborated by the Author, 2024.*

*"Emerging Advances of Blockchain Technology in Finance"*(AUTHORS, 2023c) explores the transformative role of blockchain in finance, particularly in enhancing transparency and efficiency in financial transactions. The paper examines the use of smart contracts and blockchain-based platforms for automating financial processes and ensuring compliance with regulatory standards. It concludes that blockchain can significantly improve the transparency and efficiency of financial transactions but also notes the significant regulatory and technical challenges that need to be addressed. The paper suggests a cautious but optimistic approach to adopting blockchain in finance, focusing on incremental improvements and regulatory cooperation.

*"Adoption of Blockchain Technology in Supply Chain Operations"*(AUTHORS, 2022a) discusses the increasing adoption of blockchain technology in supply chain management. It highlights blockchain's potential to improve safety, visibility, and auditability in supply chains. The paper also addresses the challenges and limitations of integrating blockchain into existing supply chain systems, concluding that blockchain can significantly improve supply chain operations by providing enhanced visibility, safety, and auditability. The study stresses the importance of addressing integration challenges and the need for a standardized approach to blockchain implementation in supply chains to maximize its benefits.

*"From Traditional Product Lifecycle Management Systems to Blockchain-Based Platforms"*(AUTHORS, 2022b) compares traditional product lifecycle management (PLM) systems with blockchain-based platforms. It argues that blockchain can overcome the inefficiencies of existing PLMs by providing a secure and connected infrastructure for data handling, processing, and storage throughout the product lifecycle. The paper concludes that blockchain can enhance PLM systems' efficiency and security, recommending its adoption to improve overall product lifecycle management.

*"Exploring Blockchain Research in Supply Chain Management: A Latent Dirichlet Allocation-Driven Systematic Review"*(AUTHORS, 2023d) uses Latent Dirichlet Allocation (LDA) to analyze blockchain research within supply chain management, identifying key themes and applications. The study provides a comprehensive overview of the current research landscape and suggests future research directions. The authors conclude that blockchain research in supply chain management is diverse, covering topics from financial transactions to traceability. They recommend that future research focus on practical implementations and long-term impacts.

*"Blockchain Technology in Supply Chain Management: Insights from Machine Learning Algorithms"*(AUTHORS, 2023a) employs machine learning algorithms to analyze blockchain research trends in supply chain management. The paper provides insights into the most prominent themes and applications of blockchain in the field. It concludes that blockchain has the potential to enhance transparency and efficiency in supply chains but highlights the need for more empirical research to validate theoretical models and address practical challenges.

*"How Blockchain Can Enhance Transparency, Traceability, and Trust in Procurement Processes"*(AUTHORS, 2023e) explores the benefits of blockchain in procurement, emphasizing its potential to provide transparency and traceability in supply chain operations. The study examines how smart contracts can automate procurement processes and enhance trust among stakeholders. It concludes that blockchain can significantly improve procurement processes by enabling transparent and traceable transactions. However, the paper also notes that addressing technical and regulatory challenges is necessary for the full realization of blockchain's potential in procurement.

| **Study Name** | **Research Scope** | **Application** | **Contribution** |
|---|---|---|---|
| A Comprehensive Review of Blockchain Technology in Supply Chain Management | Blockchain applications in supply chain management. | Supply chain management. | Enhances transparency, traceability, and trust; addresses integration, scalability, and compliance challenges. |
| Emerging Advances of Blockchain Technology in Finance | Blockchain's role in finance. | Financial transactions, smart contracts. | Improves transparency and efficiency; notes regulatory and technical challenges. |
| Adoption of Blockchain Technology in Supply Chain Operations | Increasing blockchain adoption in supply chains. | Supply chain operations. | Improves safety, visibility, and auditability; stresses standardized approach. |
| From Traditional Product Lifecycle Management Systems to Blockchain-Based Platforms | Comparing PLM systems with blockchain platforms. | Product lifecycle management. | Secure, decentralized data management, improving PLM efficiency. |
| Exploring Blockchain Research in Supply Chain Management: A Latent Dirichlet Allocation-Driven Systematic Review | Analyzing blockchain research in supply chains. | Supply chain management. | Identifies key themes, recommends focus on practical implementations. |
| Blockchain Technology in Supply Chain Management: Insights from Machine Learning Algorithms | Blockchain research trends using machine learning. | Supply chain management. | Highlights need for empirical research to validate models. |
| How Blockchain Can Enhance Transparency, Traceability, and Trust in Procurement Processes | Blockchain benefits in procurement. | Procurement processes. | Automates procurement, enhances transparency and trust; notes challenges. |

Table 5 – Summary of Blockchain Studies in Specific Applications

Source: *Elaborated by the Author, 2024.*

### 3.3.1 Differentiation from Other Studies

While most studies focus on specific applications like financial transactions, supply chain management, or SLA management, the Absolute Governance methodology provides a comprehensive framework for synchronizing and certifying the transactional states across various corporate operations. Absolute Governance emphasizes a multi-stakeholder consensus mechanism and a decentralized approach to governance, reducing the need for central authorities or external audits. It utilizes smart contracts not only for automating business rules but also for ensuring the authenticity and integrity of measurement reports and contractual obligations.

### 3.3.2 Specific Contribution of Absolute Governance

By leveraging blockchain's immutability, the Absolute Governance methodology ensures that all transactional data is tamper-proof and verifiable. The proposed framework aims to reduce the transaction costs associated with contractual measurement reports by automating data verification and validation processes. The multi-signature consensus mechanism and the use of cryptographic signatures for all actions ensure higher levels of trust and transparency among stakeholders. The framework facilitates easier compliance with regulatory requirements by providing an immutable and transparent record of all transactions and decisions. Automating business rules and processes through smart contracts reduces manual intervention, leading to significant cost savings and increased operational efficiency.

# 4 Methodology

This chapter presents the methodological framework employed to investigate the efficacy of the Absolute Governance (AG) methodology in industrial contexts. The study utilizes a mixed-method approach, integrating the technical implementation of blockchain-based tools with statistical analysis of operational data. The AG methodology serves as the method, employing blockchain technology as the primary tool to address issues of contractual disputes and overbilling in industrial transactions. The case study focuses on two industrial corporations, Steel-Supplies S.A. and Oil Drilling S.A., where the AG methodology was applied and its impact evaluated. This chapter details the research design, including a step-by-step description of the implementation process, the AG setting, and the data collection and analysis methods, providing a comprehensive understanding of the methodology that can be replicated in various industrial settings.

## 4.1 Objectives

### 4.1.1 General Objective

The general objective of this study is to evaluate the effectiveness of the Absolute Governance methodology in reducing contractual disputes, preventing overbilling, and enhancing transactional integrity within industrial sectors.

### 4.1.2 Specific Objectives

The specific objectives of this study are:

- To implement and test the AG methodology in different industrial sectors.
- To assess the impact of AG on the reduction of contractual disputes.
- To measure the effectiveness of AG in preventing overbilling.
- To analyze the improvement in transactional integrity with the use of AG.

- To compare the performance of AG with traditional contractual management methods.

## 4.2 Research Hypotheses

The study explores the following scientific hypotheses:

- A mathematically verifiable consensus mechanism managing the entire contractual relationship between two or more parties can produce tangible benefits.
- Tracking every state change using this mechanism during the contractual period reduces transactional costs.
- The decentralized automation of business rules can correct contractual payable variations, thereby reducing overbilling.

## 4.3 Research Context

This section describes the context in which the research was conducted, detailing the industrial environments, the nature of contractual processes, and the specific challenges faced by the participating companies. Steel-Supplies S.A. and Oil Drilling S.A. are large industrial corporations engaged in complex contractual relationships with multiple counterparties. These contracts involve intricate Service Level Agreements (SLAs) with detailed operational and financial terms. The challenges include frequent contractual disputes, overbilling incidents, and a lack of transparency and efficiency in transactional processes. This context is crucial for understanding the relevance and applicability of the AG methodology.

## 4.4 Case Study

The case study component of this research involves the implementation of the Absolute Governance (AG) methodology within Steel-Supplies S.A. and Oil Drilling S.A. The purpose of the case study is to observe and evaluate the real-world application of the AG methodology and its effects on contractual processes, specifically focusing on the reduction of disputes, prevention of overbilling, and enhancement of transactional integrity.

### 4.4.1 Participant Selection

Participants were selected based on the complexity of their contractual processes and their significant market presence in their respective industrial sectors. The selection criteria included:

- Significant market presence in the chosen industrial sector.
- Complex and high-stakes contractual processes.
- Willingness to engage in long-term observation and data sharing.

### 4.4.2 Observed Setting in the Current Industrial Landscape

In the current industrial landscape, companies typically synchronize Service Level Agreement (SLA) transaction states through one of two canonical modalities: unilateral or bilateral exchanges. In the unilateral modality, one company presents its version of operational reality, which the other passively accepts. This can lead to governance imbalances and potential abuses, such as information manipulation, overbilling, and data obstruction. In the bilateral modality, both companies independently prepare their respective operational truths to be merged or decided upon, which mitigates some issues but introduces new challenges, including the need for shared environments, dispute resolution criteria, and potential overbilling or litigation. These modalities are illustrated in Figure 3.

#### 4.4.2.1 Proposed Setting

The proposed Absolute Governance setting aims to address these challenges by leveraging blockchain technology to create a transparent, secure, and efficient mechanism for synchronizing transaction states. The AG methodology replaces legacy business processes by introducing a custom form or API integration for gathering measurement report data, eliminating the need for separate documents sent via email. Spreadsheet or ERP modules used for SLA calculations, which are prone to formula alterations and

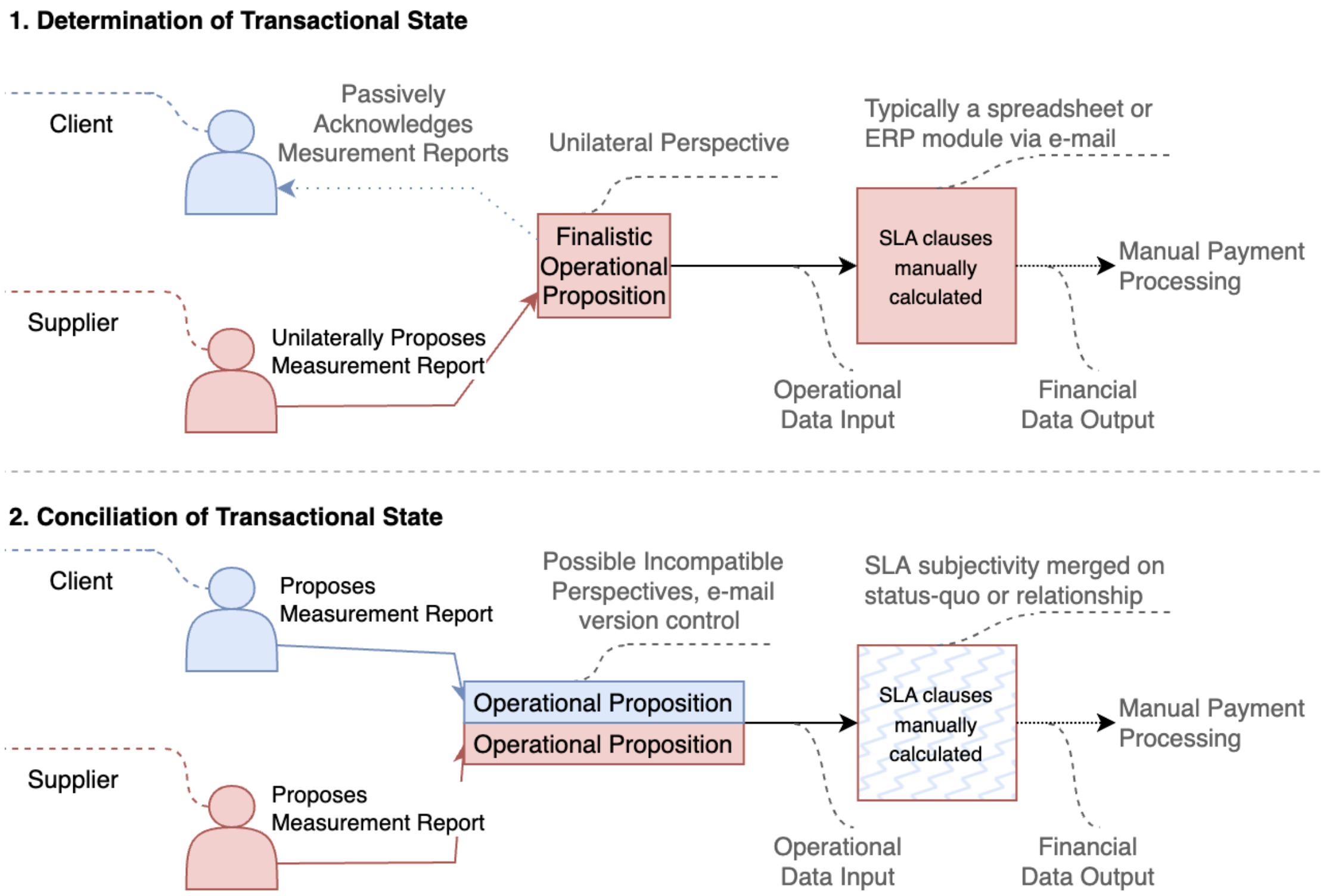

Figure 3 – Current Setting: The canonical pathways for synchronizing transaction state without blockchain technology.

Source: *Elaborated by the Author, 2024.*

human errors, are replaced with procedurally automated calculations via SLA smart contracts, as demonstrated in Figure 4.

This methodological update enables the enhancement of the state synchronization pipeline by integrating dispute resolution mechanisms such as arbitration chambers or consulting firms to mediate consensus. Beyond automating payable calculations, the payment itself could be automated via interaction with a Central Bank Digital Currency (CBDC)-compatible smart contract. These and other scenarios are discussed in more detail in Chapter 5.

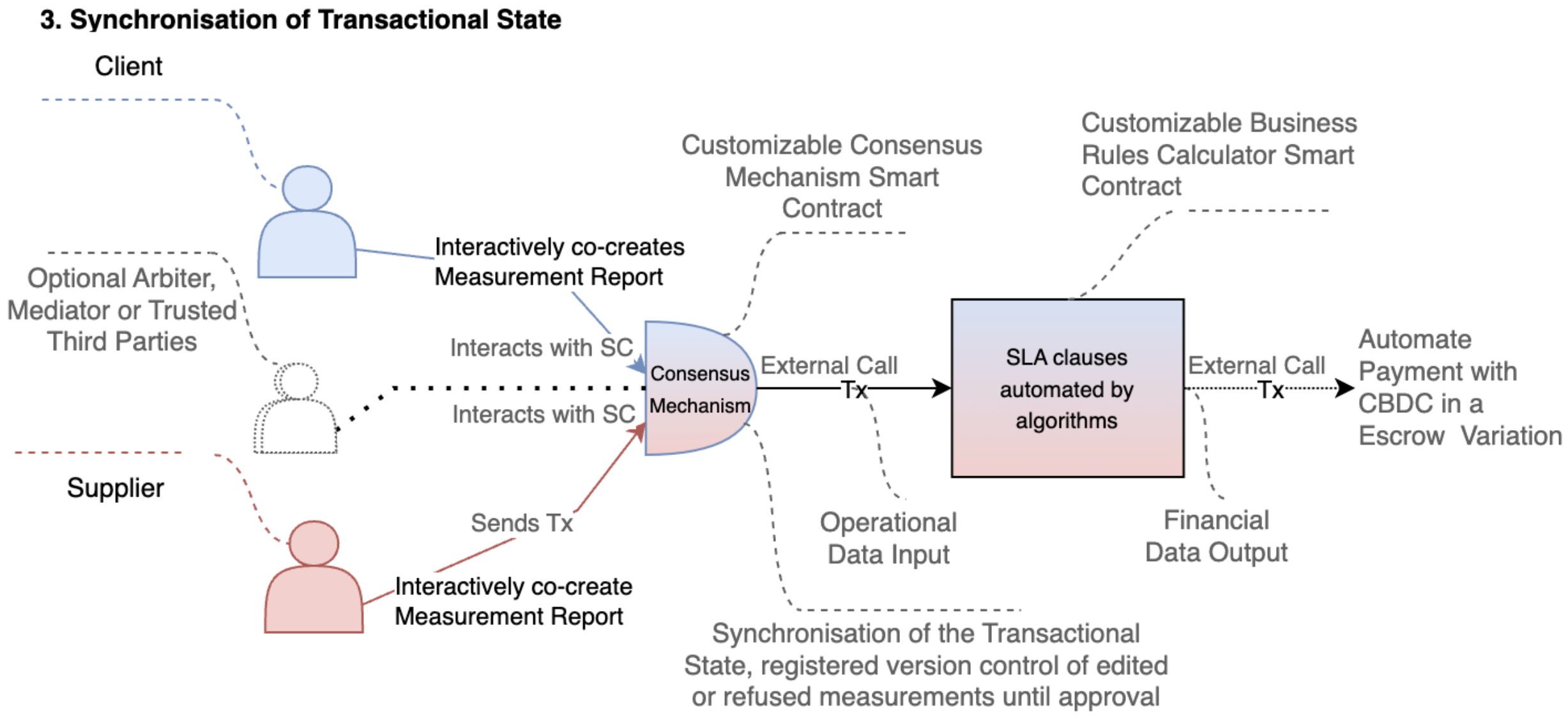

Figure 4 – Proposed Setting: The emerging standard for synchronizing transaction state based on blockchain technology.

Source: *Elaborated by the Author, 2024.*

## 4.5 Design of the Proposed Solution

The general architecture of the Absolute Governance methodology implementation consists of technical artifacts designed to ensure the following premises:

1. **Participant Identification**: Every participant is identified by an Ethereum address, known only to them, with roles assigned according to the governance rules.

2. **Consensus Mechanism**: A customized consensus mechanism adheres to the governance rules of the contract, allowing operational data to be inputted, evaluated, edited, and approved until a definitive consensus is reached.

3. **Operational to Financial Transformation**: The transformation of operational state into financial state is executed procedurally in a decentralized and auditable manner, eliminating the possibility of human intervention or fraud.

### 4.5.1 Design of the Smart Contracts Architecture

The solution design relies on smart contracts to assert non-repudiation, data integrity, and streamlined calculation automation. A platform with a user interface was developed for users to be assigned Ethereum addresses, while solely controlling their private keys via ERC-4337 account abstraction (QUICKNODE, 2023). Each address receives a Non-Fungible Token (NFT) representing their roles and voting power according to their responsibilities.

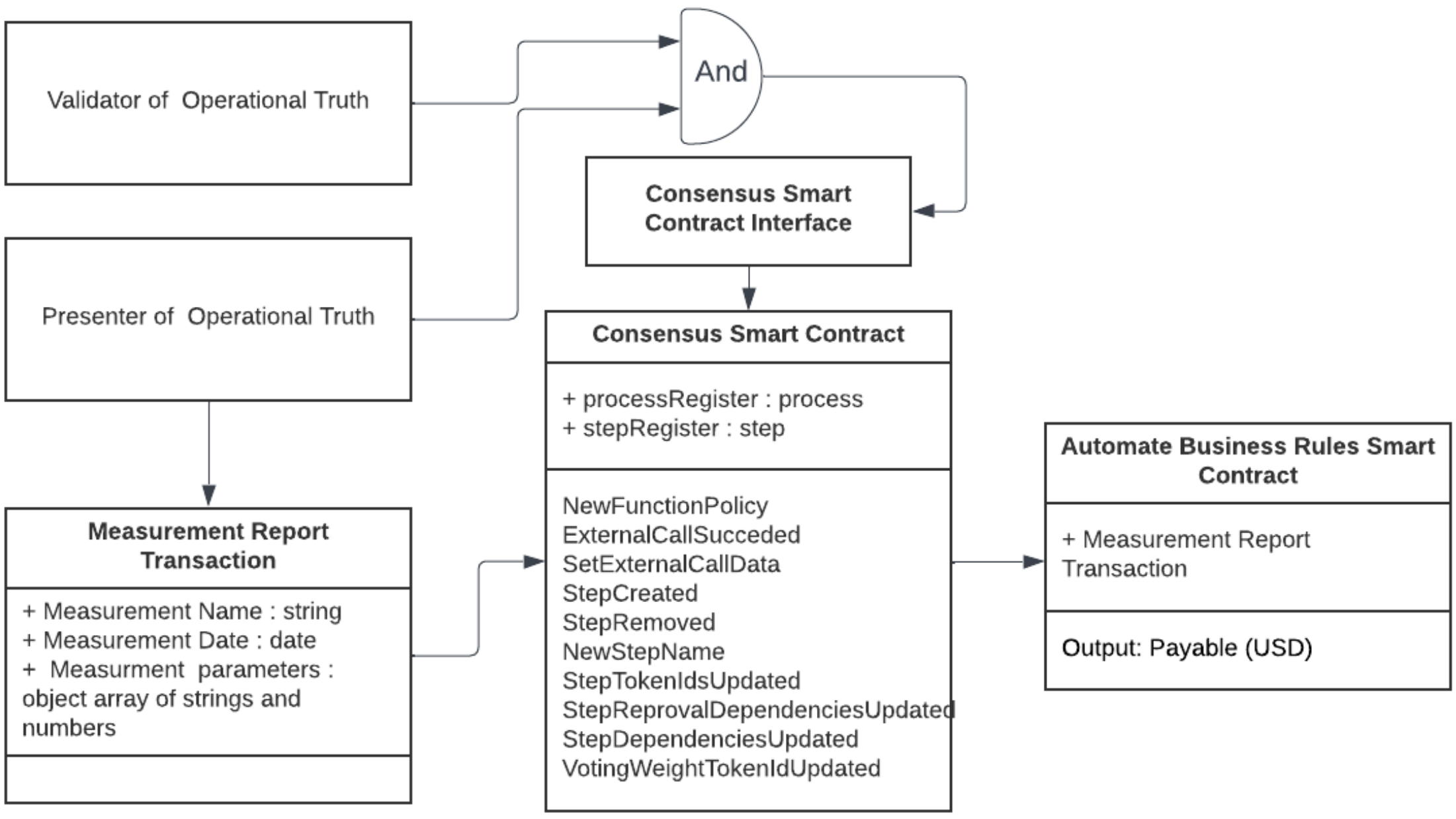

Figure 5 – UML diagram illustrating the implementation of smart contracts in the Absolute Governance methodology.

Source: *Elaborated by the Author, 2024.*

To transform the operational data into financial data procedurally, a calculation smart contract with an algebraic library is deployed to represent all the quantitative SLA business rules from the traditional contract in a transparent and auditable form. High-level smart contract schematics can be seen in Figure 5.

This configuration ensures that all propositions and alterations of the transactional state are voluntarily agreed upon, signed, and registered for future consultation.

Simultaneously, given an operational truth, the smart contract automatically converts it into an uncontroversial payable outcome.

### 4.5.2 Design of the Consensus Mechanism Smart Contract Architecture

Understanding the dynamics of the consensus mechanism smart contract is essential to comprehend the AG design implementation choices. This contract organizes the version control of the proposed operational state with functionalities to track and register all events, from proposition to approval, rejection, and intermediary versions of the measurement reports.

Each contractual participant can propose a version of the operational state. Two actions are triggered: first, all operational data is concatenated, hashed, and stored in a public blockchain with a high Nakamoto coefficient (METRICS, 2020). Actions are registered through contract interactions, and report data via SHA256 hash functions, ensuring verifiable data integrity while preserving full information security and protection of industrial secrets (DILHARA, 2020).

An interaction with the smart contract creates a pending proposal review on the state machine. The party receiving the proposed operational truth can read and make observations contributing to the document's cohesion, which can then be approved or rejected.

If approved, a final consensus is reached, and a timeline and certificate are issued. If rejected, the proposed state can be edited and resubmitted for as many revision cycles as necessary until approval is achieved, as demonstrated in Figure 6. The approval criteria are configurable, with adaptable rules to accommodate a different number of participants and distinct weights for each.

## 4.6 Implementation Phases

The AG methodology was structured into three distinct phases, designed to systematically implement and evaluate the methodology within the participating industrial corporations.

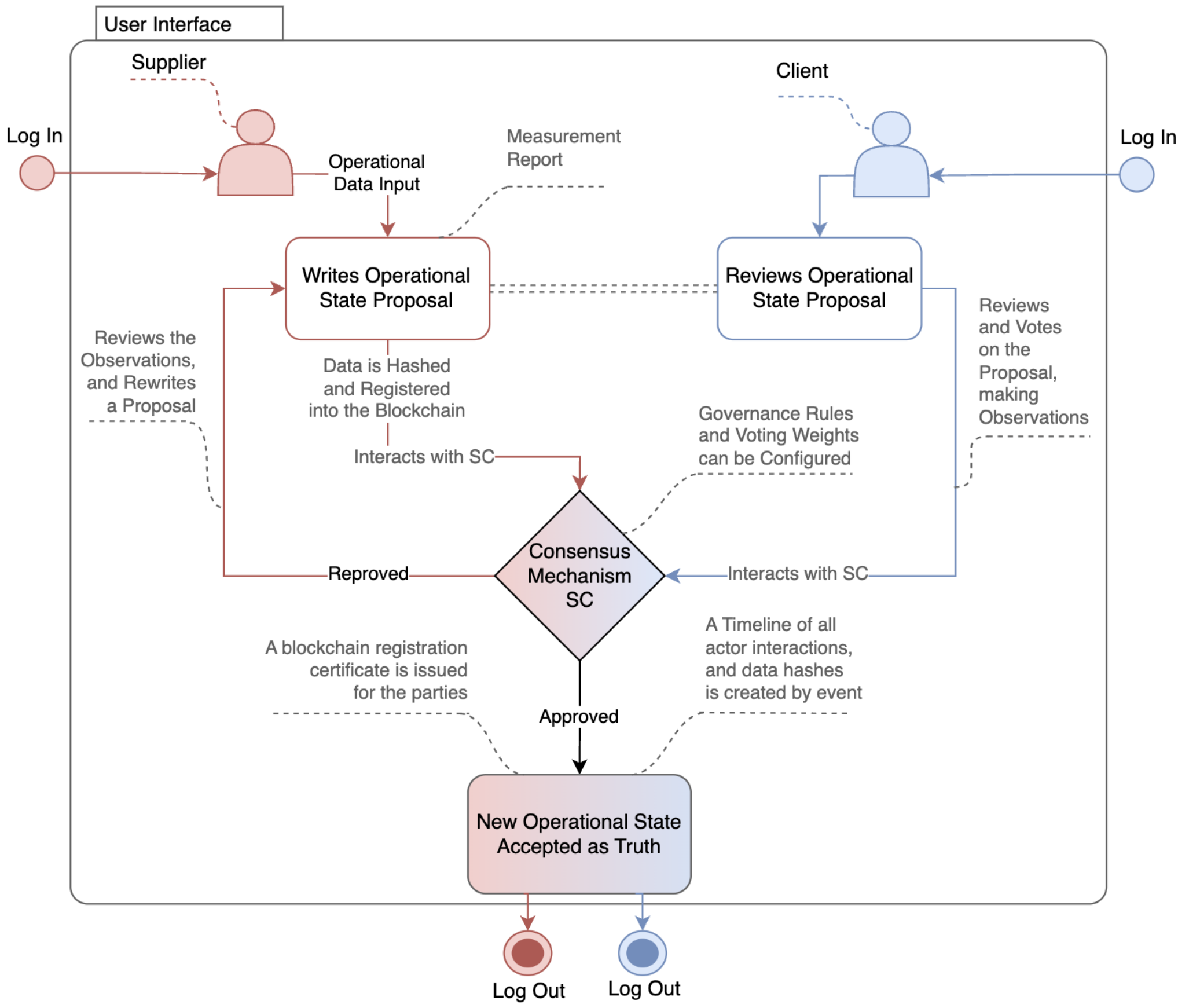

Figure 6 – UML diagram illustrating the Consensus Mechanism Smart Contract High-Level Architecture

Source: *Elaborated by the Author, 2024.*

#### 4.6.0.1 Phase 1: Digitalization of Measurement Report Processes

The initial phase involved the digitalization of existing measurement report processes using a no-code BPMN (Business Process Model and Notation) platform. This platform allowed for the mapping and definition of data names and types necessary to digitize the current processes. Instead of sending measurement reports via email, the information became available through a consensus mechanism, ensuring data integrity and transparency. The steps in this phase included:

1. **Process Mapping**: Identifying and documenting existing measurement report processes.

2. **Data Modeling**: Defining the data structures required for digital representation.
3. **Platform Configuration**: Setting up the BPMN platform to replicate the processes digitally.
4. **User Training**: Educating staff on how to use the new digital tools.

#### 4.6.0.2 Phase 2: Development of the SLA Smart Contract

In this phase, all qualitative SLA rules were transformed into programmable algorithms within smart contracts. The smart contracts underwent rigorous testing to verify functionality, security, and compliance before deployment on the blockchain network. Integration with existing IT systems was also established to ensure seamless data transmission and recording. Steps included:

1. **Rule Extraction**: Identifying and extracting all SLA business rules from traditional contracts.
2. **Algorithm Development**: Translating business rules into programmable algorithms.
3. **Smart Contract Coding**: Developing smart contracts using Solidity or other appropriate languages.
4. **Testing and Validation**: Performing unit tests, security audits, and compliance checks.
5. **System Integration**: Linking smart contracts with existing ERP systems and databases.

#### 4.6.0.3 Phase 3: Implementation and Measurement

The final phase involved deploying the blockchain environment tailored to the company's specific needs and conducting continuous monitoring and data collection. Key steps included:

1. **Environment Deployment**: Setting up the blockchain network and smart contracts for live operation.

2. **User Interaction**: Facilitating user interaction with data sources and ensuring ease of use.

3. **Monitoring**: Continuously tracking transactions, smart contract executions, and system performance.

4. **Data Collection**: Gathering data on contractual measurement reports, overbilling incidents, and disputes.

5. **Analysis Preparation**: Preparing the collected data for statistical analysis.

## 4.7 Data Collection and Analysis

Data collection focused on capturing both operational and financial aspects of the contractual processes. Specifically, the following data were collected:

- **Measurement Reports**: Detailed records of each measurement report processed through the AG methodology.
- **Blockchain Transactions**: Logs of all blockchain transactions, including timestamps, participants, and transaction hashes.
- **Contractual Disputes**: Records of any disputes arising during the contractual period, including nature and resolution.
- **Overbilling Incidents**: Instances where overbilling was identified, quantified, and corrected through the AG methodology.
- **Traditional Calculations**: Parallel data from traditional contractual management methods for comparative analysis.

For analysis, the data were subjected to statistical tests, including Analysis of Variance (ANOVA), to assess the impact of the AG methodology on overbilling reduction and transactional efficiency. Comparative analyses were conducted between the AG methodology and traditional methods to establish statistical significance and validate the research hypotheses.

# 5 Results

## 5.1 Study Duration and Period

This chapter presents the empirical findings and analytical insights derived from the application of the Absolute Governance (AG) methodology within two major industrial corporations: Steel-Supplies S.A. and Oil Drilling S.A. The chapter is structured with the duration of the study, the comparative analysis of results across the observation period, and the statistical validation of the impact of blockchain technology on contractual processes. Furthermore, it discusses the practical implications of these findings, the broader significance of blockchain integration in industrial operations, and outlines the prerequisites and scalability potential for the implementation of AG in other contexts.

### 5.1.1 Participants

The study primarily focused on two large industrial corporations, anonymized as Steel-Supplies S.A and Oil Drilling S.A. These entities were chosen due to their significant market presence and the complex nature of their contractual processes.

### 5.1.2 Study Duration and Period

This case study was conducted from 2019 to 2023. The timeframe allowed for the collection and analysis of monthly measurement report data and for a detailed assessment of the long-term impacts of implementing a blockchain-based Absolute Governance methodology within complex industrial sectors.

For Steel-Supplies S.A., the observation period spanned from 2019 to 2023, during which the company engaged with five distinct industrial counterparties. Throughout this period, 26 contractual measurement reports were meticulously analyzed, involving 78 blockchain transactions, each authenticated with blockchain signatures.

As of Oil Drilling S.A., the study covered the years 2022 and 2023. The interactions of this company were limited to one primary industrial counterparty. The data collected included eight detailed contractual measurement reports, encompassing 32 blockchain transactions and signatures.

#### 5.1.2.1 Comparison Period

The comparison period for both industries was the same as the study period. Methodologically, SLA measurements were performed redundantly. At the same period, the legacy calculations were performed alongside the AG smart contract calculations, while being compared monthly.

The application of the Absolute Governance (AG) methodology, which consists of a corporate consensus mechanism on the operational state followed by a smart contract, as illustrated in figure 4, within Steel-Supplies S.A. and Oil Drilling S.A., spanning from 2019 to 2023, demonstrated significant empirical outcomes in TOC reduction. The methodology is an In Steel-Supplies S.A, across a two-year implementation phase (2021-2023), with five industrial counterparties, there were 26 contractual measurement reports in which the last three were fully automated. A notable reduction of 10% in contractual disputes was observed, alongside a measurable decrease in the measurement report calculation time, which impacted the invoicing anticipation process by an average of 5 days. Most importantly, contract miscalculations, including overbilling, were reduced by 2.42%.

In the context of Oil Drilling S.A, during the shorter span of 2022 to 2023 and engaging with a single counterparty, 8 contractual measurement reports were analyzed. The company experienced a 20% reduction in contractual disputes, an impressive average reduction of 14 days in the measurement report calculation time, and a 1.8% decrease in contract miscalculations leading to direct savings.

By comparison of the contractual dispute rates from previous periods, an enhancement on contractual fluidity attributed to the synchronization of the operational state via a consensus mechanism with decentralized records can be perceived on both subjects, because dispute data is scarce and sporadic no statistical analysis was conducted in order to prevent bias.

As can be observed in figure 5.1.2.1, all automated measurement reports presented some level of overbilling. The most common of the speculative, but not extensive, causes of overbilling were, non application of contractual clauses, rounding, human

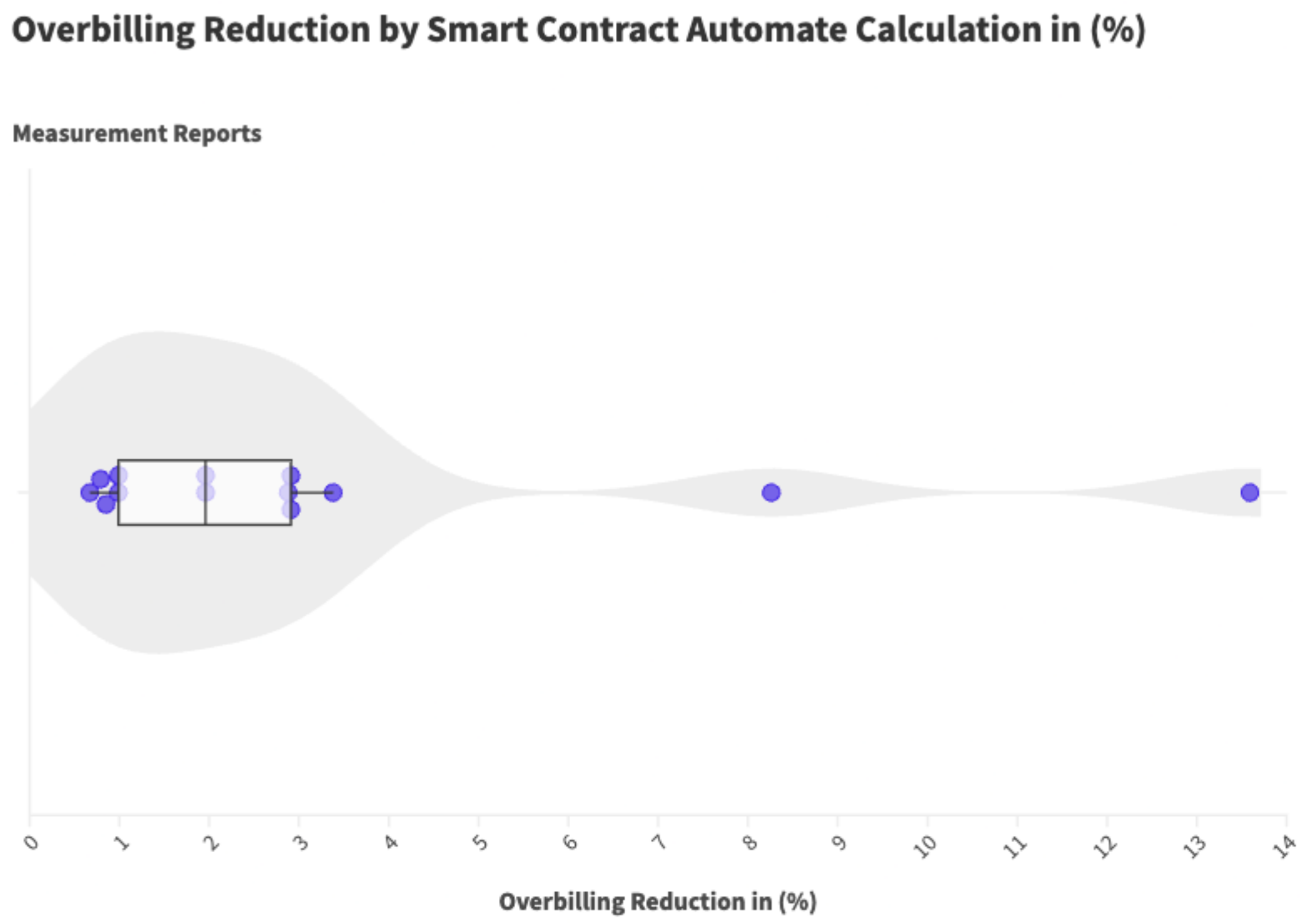

Figure 7 – Industrial Measurement Reports Automation from 2021 to 2023.

Source: *Elaborated by the Author, 2024.*

error, and traditionalism on status-quo. Taking into account the contract calculation data, the SLAs studied presented a fixed and recurrent measurement period of monthly cycles, which enabled a deeper quantitative analysis. From a statistical standpoint, the study employed an ANOVA analysis to scrutinize the overbilling reduction, revealing an F-ratio value of 9.98155 and a p-value of 0.004237. The statistical significance of these results, at $p < 0.01$, robustly rejects the null hypothesis, establishing that the algorithmic implementation of business rules automation via smart contracts led to a measurable and substantial decrease in overbilling incidents.

## 5.2 Discussion

The integration of blockchain technology, particularly the AG methodology, into the contractual processes of these industrial giants demonstrated not only a reduction in contractual disputes but a reduction in overbilling. A robust indicative of the visibility and conformity that blockchain and smart contracts introduce into traditional business processes.

From an economic standpoint, the reduction in receivables anticipation days signifies an improvement in cash flow management, a crucial aspect for any industrial operation. The digital methodology to standardize the management of operational reports and occurrences, and the automation provided by smart contracts, directly contribute to these efficiencies, mitigating the risk of human error and subjective interpretation of contractual terms.

The study highlights the robustness of blockchain as a secure and transparent medium for contractual agreements. The cryptographic nature of blockchain provides a level of security and immutability that traditional digital transactions cannot match. This security is paramount in industries where the stakes of contractual breaches or disputes are high.

### 5.2.1 The Need for a Blockchain Application in State Synchronization

The difference of non-blockchain solutions and the Absolute Governance methodology, permeate operational efficiency, information security, contract governance, and ESG. Companies invests in storing and managing the operational state, but in times of crisis, centralized security information can be compromised as stated in section 1.2. The use of a private consensus mechanism application with a blockchain integration serves as trust layer for contractual state synchronization, which beyond preserving data integrity, solves a legacy problem of lack of visibility of contract conformity and operational diligence. The lack of visibility fosters subjectivity and unpredictability, leading rational players to increase their margins to accommodate operational risks, therefore increasing transactional costs on the supply chain. Non-blockchain solutions are always editable by the owner, leading to the impossibility of change history verification by the counterparty; by solving this dilemma, contractual disputes are positively impacted as stated in chapter 5. Furthermore, instead of relying on hired auditing reports, from auditing and rating companies, which can be compromised by agency problems (WATTS; ZIMMERMAN, 1983), the stakeholders of the firm can, in an unprecedented way, draw information from a source with verifiable integrity, which was validated by a counterparty that has diverse, if not diametrical, interests such as in the Supplier-Client game. The legitimacy of the "skin in the game"(TALEB, 2018) of the data validation from the counter-party immersed in a transaction presents a model of corporate state verification alternative to the reputation lending of the audit reports from third party agencies. The possible granularity and extent of transaction verification by counterparties using the consensus mechanism can provide unbiased reality reports that largely upgrade conventional contract and processes governance routines, thus improving ESG scores. (KALEIDO, 2023) This presented methodology of establishing governance, without necessarily relying on external parties alien to the transaction, promotes trough blockchain technology an institutional Transactionalty that is cost-efficient and replicable.

### 5.2.2 Prerequisites for Implementation

Although all processes and contracts related to state synchronization are suitable for receiving the implementation of the AG, the costs associated with any technology incentivize prioritization. The anatomy of good candidates for the methodology presents the following characteristics:

- The party and counterparties understand transaction costs from lack of contract visibility and conformity, and are willing to establish a work ethic ridden of subjectivity.
- The chosen measurement report process is recurrent, with at least one measurement per year.
- The quantitative business rules of the SLA are minimally stable. Although prices and conditions may be configured as mutable variables, the structure of the business rules that represent the calculation functions should not change frequently.

Empirically, if all three points are properly observed, the project implementation incurs in a significant risk reduction.

### 5.2.3 Generalization and Scaling

To explore scalability, the technological cost structure of the AG solution should be stated. The main two cost structures are recurrent costs and implementation costs.

Just like any cloud solution, the platform with user interface scales with usage regarding infrastructure and server costs, the main resources consumed are relational database and serverless event driven computer services. The blockchain transaction costs are negligible in most scenarios, considering when business process associated with measurement reports only presents macro recurrences of daily, weekly, or monthly timeframes. Regarding implementation costs, any process can be configured by the user in a no-code platform using a BPMN library in the same manner a form is set to be fulfilled by the end users. This self-service configuration workflow supports a user-friendly and intuitive mapping of analog process to digital. The scalability bottleneck lies in configuring domain-specific SLA rules, which requires substantial development effort. This challenge is addressed by employing a general calculation library, pre-fabricated SLA templates, and artificial intelligence code generation tools to streamline the configuration process. Given this panorama of the cost structure, materials and inputs, the experience of the implementors in scaling the solution within diverse industries is promising. In the roadmap of implementation, the total time invested diminished substantially comparing the first and second instances, reaching intramonth intervals. The present results of AG can be generalized with the deployment of the technology in several areas of the same company and in completely different industries without specific knowledge requirements, since all processes and contracts are reduced to BPMN and discrete algebraic algorithms. Parallel implementations are absolutely possible with the correct development resource allocation, focusing on customization.

# 6 Conclusion

## 6.1 Concluding Remarks

The results obtained from the application of the Absolute Governance methodology within Steel-Supplies S.A. and Oil Drilling S.A. align closely with the core and specific objectives outlined in this dissertation. The primary objective of this research was to establish a blockchain-based solution for the classical problem of contractual synchronization between institutions, which was achieved by implementing by a consensus mechanims of the measurement reports of a SLA contract. The notable reduction in contractual disputes, overbilling incidents, and improvement in invoicing processes directly support the hypothesis that a mathematically verifiable consensus mechanism can reduce transactional costs and enhance transactional integrity. Specifically, the construction and application of the AG methodology, as well as the design of the consensus mechanism architecture, were validated through empirical evidence from the case studies. The reduction in contractual disputes by 10% and 20% in Steel-Supplies S.A. and Oil Drilling S.A. respectively, confirms the effectiveness of the methodology in fostering contractual truthfulness and mitigating transactional uncertainties. The statistical analysis further reinforced the significance of the results, demonstrating that the automation of business rules through smart contracts can lead to substantial cost savings and operational efficiencies.

These outcomes not only demonstrate the practical applicability of the proposed solution in real industrial environments but also highlight the potential for scalability and generalization of the AG methodology across different sectors. The findings affirm the dissertation's objectives of promoting a resilient, transparent mechanism for validating transactional states and establishing Absolute Governance as a powerful tool for improving corporate governance and operational efficiency in the context of Industry 4.0. The study corroborates the critical significance of blockchain technology in streamlining contractual governance within industrial sectors in order to reduce total operating costs. The Absolute Governance methodology implementation yielded quantifiable improvements in overbilling reduction (average = 2.11%). These results not only reduced transaction costs but pave the way for establishing a new governance standard where companies are capable of synchronizing transactional state in order to consensually automate business rules calculations to reduce transaction costs.

## 6.2 Future Work

There are distinct research lines to be studied on the basis of the AG methodology. These can be classified in the further technical developments and theoretical advancements. The first refers to the implementation of new features in the consensus mechanism, industry-specific arbitrage or mediation chamber can be integrated to add a nonjudicial, cost-effective and fast alternative to the dispute resolution process. The other rational improvement from the automatic calculation of the SLA payable is the integration with CBDCs to also streamline direct payments. To this functionality to work properly, the consensus mechanism smart contract should be upgraded with the capacity to receive funds, becoming an escrow contract capable of routing payments back and forward based on governance decisions.

The second research line to follow is the application of AG in other industries that heavily rely on SLA contracts. This future work proposition can determine whether the methodology has the potential to become a market standard.

# Remissive Index

## T

## U

## V